\documentclass[twocolumn,iop]{aastex61}
\pdfoutput=1 
\usepackage{amsmath,amstext}
\usepackage[T1]{fontenc}
\usepackage{apjfonts} 
\usepackage{hyperref}
\usepackage[figure,figure*]{hypcap}

\shorttitle{Clustered Star Formation in the Galactic Disk}

\shortauthors{Kamdar et al.}

\begin{document}

\title{A Dynamical Model for Clustered Star Formation in the Galactic Disk}

\author{Harshil Kamdar\altaffilmark{1}, Charlie Conroy\altaffilmark{1}, 
Yuan-Sen Ting\altaffilmark{2,3,4,5}, Ana Bonaca\altaffilmark{1}, Benjamin Johnson \altaffilmark{1}, Phillip Cargile\altaffilmark{1}}

\altaffiltext{1}{Harvard-Smithsonian Center for Astrophysics, 
Cambridge, MA, 02138, USA} 
\altaffiltext{2}{Institute for Advanced Study, Princeton, NJ 08540, USA} 
\altaffiltext{3}{Department of Astrophysical Sciences, Princeton University, Princeton, NJ 08544, USA} 
\altaffiltext{4}{Observatories  of  the  Carnegie  Institution  of  Washington,  813  Santa Barbara Street, Pasadena, CA 91101, USA} 
\altaffiltext{5}{Hubble Fellow}

\begin{abstract}
The clustered nature of star formation should produce a high degree of structure in the combined phase and chemical space in the Galactic disk. To date, observed structure of this kind has been mostly limited to bound clusters and moving groups. In this paper we present a new dynamical model of the Galactic disk that takes into account the clustered nature of star formation. This model predicts that the combined phase and chemical space is rich in substructure, and that this structure is sensitive to both the precise nature of clustered star formation and the large-scale properties of the Galaxy. The model self-consistently evolves 4 billion stars over the last 5 Gyr in a realistic potential that includes an axisymmetric component, a bar, spiral arms, and giant molecular clouds (GMCs). All stars are born in clusters with an observationally-motivated range of initial conditions. As direct \textit{N}-body calculations for billions of stars is computationally infeasible, we have developed a method of initializing star cluster particles to mimic the effects of direct \textit{N}-body effects, while the actual orbit integrations are treated as test particles within the analytic potential. We demonstrate that the combination of chemical and phase space information is much more effective at identifying truly co-natal populations than either chemical or phase space alone. Furthermore, we show that co-moving pairs of stars are very likely to be co-natal if their velocity separation is $< 2$ km s$^{-1}$ and their metallicity separation is $< 0.05$ dex. The results presented here bode well for harnessing the synergies between \textit{Gaia} and spectroscopic surveys to reveal the assembly history of the Galactic disk.

\end{abstract}

\keywords{Galaxy: evolution -- Galaxy: kinematics and dynamics -- open clusters and associations: general}
\section{Introduction}

The overarching goal of the field of Galactic archeology is to reconstruct the formation history of the Milky Way based on observations of stars in our Galaxy today \citep[e.g.,][]{eggen1974oldest, freeman2002new, schonrich2009chemical, bland2010long, mitschang2014quantitative, de2015galah}. Galactic archeology aims to address a number of topics including, but not limited to, the star formation history of the Milky Way, chemical evolution, studying the first generation of stars in the Milky Way, and the dynamical history of the Galaxy. In the context of the Galactic disk, the building blocks of star formation -- star clusters -- enable us to probe the chemodynamical evolution of the Milky Way \citep[e.g.,][]{lada2003embedded, goodwin2006gas, bastian2006evidence, portegies2010young, longmore2014formation, walker2015tracing}. However, due to the short dynamical time in the disk, unbound star clusters are phase mixed quickly, making their study difficult. 

The traditional view of star cluster evolution \citep[e.g.,][]{lada2003embedded, fall2005age, goodwin2006gas, 2008MNRAS.384.1231B, assmann2011popping} argues that all or most stars are born in star clusters but most disrupt on short timescales. \citet{lada2003embedded} provided strong evidence for this scenario by finding embedded, gas-rich star clusters ten times as often as gas-free open clusters in the Galaxy. Making the assumption that embedded clusters are ancestors of classical open clusters, the dearth of the latter led \citet{lada2003embedded} to suggest that a small fraction ($\lesssim 10$\%) of bound star clusters are able to survive the embedded phase due to gas expulsion. The unbinding of young star clusters due to gas expulsion was dubbed "infant mortality". 

However, recent observations \citep[e.g.,][]{bressert2010spatial, gutermuth2011correlation, parker2012characterizing, kuhn2014spatial, grasha2017hierarchical, elmegreen2018dispersal} of the Milky Way and other galaxies show a hierarchical, spatially correlated distribution of star-forming regions.  \citet{kruijssen2012fraction} and others argue that these results imply that surface density thresholds used to make the classical "infant mortality" argument are arbitrary and do not correspond to a physical scale. Consequently, they argued that the scarcity of bound gas-free clusters is simply a consequence of the hierarchical nature of the ISM: only a small fraction of star formation reaches densities high enough to result in bound star clusters. These two different formation channels (one where there is a lower limit to the mass of a cluster and the other where star formation is a spatially correlated, hierarchical process with practically no lower mass limit) might be directly testable through data from the \textit{Gaia} mission \citep{brown2018gaia}. Understanding the extent to which phase space and chemical information of the Galactic disk can discriminate between these two scenarios is one of the motivations for the present work. 

The study of structure on larger spatial scales has been extensively studied through both recent \textit{N}-body simulations \citep[e.g.,][]{dehnen2000effect, quillen2005effect, minchev2010new, hunt20184, laporte2018response, khoperskov2018echo} and \textit{Gaia} DR2 observations \citep[e.g.,][]{katz2018gaia, antoja2018dynamically, poggio2018galactic}. These works have shown a significant amount of structure in the solar neighborhood that has been mostly attributed to resonances of the spiral arms and/or bar and the influence of the Milky Way's satellites on the disk. However, the question of how much structure we expect to observe due to clustered star formation in the Milky Way has not been investigated in detail. 

Due to the absence of an easily noticeable phase space signature of disrupting star clusters beyond at most a couple of Galactic dynamical times, \citet{freeman2002new} proposed the idea of chemical tagging: using the chemistry of stars to understand the assembly history of the Milky Way \citep{bland2010long}. The fundamental assumption of chemical tagging is that stars born in the same star cluster will have identical initial element abundance patterns \citep[e.g.,][]{friel2002metallicities, de2007chemically, koposov2008automated, de2009reconstructing, feng2014early,mitschang2014quantitative, ting2015prospects, anders2018dissecting}. The intrinsic dispersion in the chemistry of stars born together is thought to be very low and consistent with the measurement uncertainty \citep[$<0.03$ dex; e.g.,][]{bovy2016chemical, ness2018galactic}. Therefore, in principle, it should be possible to reconstruct disrupted star clusters through their unique chemical tags.

\citet{ting2015prospects} (hereafter T15) used a flexible model of the chemical space spanned by disk stars of the Galaxy to show that, in practice, a given clump in chemical space is likely not a strictly co-natal population. T15 found that the most important factors limiting the identification of disrupted clusters via chemical tagging are the precision of elemental abundances, the dimensionality of chemical space, and the survey sampling rate of the underlying stellar population. In their fiducial model, T15 found that the dominant star cluster for a clump in chemical space only contributed 25\% of the stars in that clump. This is a rather significant problem for strong chemical tagging -- finding a set of stars clustered in chemical space does not necessarily mean that they were born together. Relatedly, \citet{ness2018galactic} found similar contamination from field stars in chemical space. 

As a result, neither phase space nor chemical space alone have had much success to date in unraveling the assembly history of the Milky Way disk. However, recent results \citep{2019A&A...621L...2R, meingast2018extended, meingast2019extended} hint at the possibility of rich substructure in the disk due to clustered star formation. The possibility of uncovering the assembly history of the disk by analyzing substructure motivates the work presented here. This paper is the first in a series that develops a new suite of models for the full population of star clusters comprising the Galactic disk. One of the primary goals of this project is to study the chemo-dynamical information content of the Milky Way disk in the \textit{Gaia} era. 

The paper is structured as follows. Section \ref{sec:methods} discusses the different components of the model in detail. Section \ref{sec:global} show global properties of the simulated galaxy compared to the Milky Way. Section \ref{sec:kinematic_chemical} combines kinematic and chemical information in the context of chemical tagging and co-moving pairs. Section \ref{sec:discussion} outlines the limitations of the model presented in this work and potential directions for future work. Section \ref{sec:summary} is a summary of the key results and assumptions made in the model. 

\section{Methods}
\label{sec:methods}
\subsection{Overview} 

We simulate the dynamical evolution of the galaxy with orbit integration of test particles coupled to a semi-analytical model of initializing the clusters in a way that roughly approximates their evolution in the disk in the backdrop of a realistic galactic potential. The key model ingredients are a star formation history, a potential for the Milky Way, a computationally feasible model for how stars in star clusters dissolve, a chemical evolution model, and orbit integration of test particles.  

This section describes the model in detail. Throughout this section, we will refer to Figure \ref{fig:model}, which shows various important model ingredients. First, we discuss the various components of the potential of the Milky Way that are implemented in our simulations (Section \ref{sec:galaxy_pot}). Second, we examine the star formation history adopted for the simulations and the radial growth of the simulated galaxy over time (Section \ref{sec:sfh}). This tells us when, where, and how many stars are born in the simulated galaxy. Third, we describe how clustered star formation is implemented in our simulations and introduce a subgrid model to mimic cluster disruption (Section \ref{sec:sc}). Fourth, we discuss the semi-empirical chemical model for the simulation, which assigns a chemical tag to every star based on when and where they were born (Section \ref{sec:chemistry}). Fifth, we explain the need for a smooth background model to represent older stars in the simulation (Section \ref{sec:bg}). Sixth, we discuss how we calculate the photometry for stars in the simulation and create our mock catalogs to enable a fair comparison with \textit{Gaia} DR2 (Section \ref{sc:phot}). Seventh, we discuss how we calculate phase space densities (Section \ref{sec:pd}). Finally, we end this section with a description of the different variations of the simulations that we have run (Section \ref{sec:variation}).

\subsection{Galaxy Potential}
\label{sec:galaxy_pot}

The galaxy modeled in this work has six components: the disk, the bulge, the bar, the halo, spiral arms, and giant molecular clouds. The form of the potential and the parameters chosen are largely based on previous work \citep{pichardo2004models, jilkova2012origin, martinez2014radial, gustafsson2016gravitational}. Table 1 lists the most important parameters of each different component and the following subsections discuss the meaning of these parameters. 

\subsubsection{Axisymmetric Component} 

The disk is modeled with a Miyamoto-Nagai disk, the expression for which is given by: 

\begin{equation}
\Phi(R, z) = -\frac{GM_d(t)}{\sqrt{R^2 + (a_{disk} + \sqrt{z^2 + b_{disk}^2})^2}}.
\end{equation}
\noindent
The parameters $M_d, a_{disk}, b_{disk}$ are given in Table 1. The $M_d$ value is the value of the mass of the disk at $z=0$; however, we expect that the disk has grown over time. Consequently, the value for $M_d$ tracks the evolution of the accumulated stellar mass by integrating the SFH as shown in Figure \ref{fig:model} (bottom right panel). The radial growth model that we adopt for the disk (Eq. \ref{eq:growth}) implies that the scale length of the disk will have changed negligibly in the last 5 Gyr as the accumulated stellar mass in the disk also changes only slightly. Consequently, we fix $a_{disk}$ to a constant value. 

The bulge is modeled with a simple Plummer sphere. The potential of a Plummer sphere is given by: 

\begin{equation}
\Phi(r) = -\frac{GM_b}{\sqrt{R^2 + b_{bulge}^2}}.
\end{equation}
\noindent
The parameters $M_b$ and $b_{bulge}$ are given in Table 1. 

The dark matter halo is modeled with an NFW profile. The potential for this model is given by: 
\begin{eqnarray}
\Phi(r) = -\frac{M_h}{r} \ln \left(1 + \frac{r}{a_{halo}}\right).
\end{eqnarray}
\noindent
The parameters used for the halo are listed in Table 1. The halo is assumed to be static and not evolving since particles are integrated in our simulation for 5 Gyr and the mass of the halo is not expected to have undergone a large change in that time \citep{behroozi2013average}.  

\subsubsection{Spiral Arms}

The spiral arms of a Milky Way-like galaxy play an essential role in the dynamical evolution of the Galaxy through processes such as radial migration and in-plane heating \citep[e.g.,][]{sellwood2002radial, minchev2010new, vera2014effect, grand2016radial}. The exact origin and persistence of spiral arms is still contested, with some arguing for a superposition of transient co-rotating structures which wind up and disappear on short timescales \citep[e.g.,][]{baba2013dynamics, sellwood2014transient} and others arguing for more long-lived modes \citep[e.g.,][]{d2013self, grand2015spiral}. In simulations, spiral arms are usually represented as periodic perturbations of the axisymmetric potential. We use the prescription given by \citet{cox2002analytical}, which models such perturbations in three-dimensional space and calculates the effect of this potential in a rotating frame. The potential of the spiral arms is given by the following expression:

\begin{equation}
\begin{split}
\Phi_\mathrm{sp}=& -4\pi GHA_\mathrm{sp}\exp{\left(-\frac{r_\mathrm{rot}}{R_\Sigma}
\right)}\sum\limits_n\left(\frac{C_n}{K_nD_n}\right)\times \\
&\cos({n\gamma}) \left[\mathrm{sech}\left( \frac{K_nz_\mathrm{rot}}{\beta_n}\right) \right]^{\beta_n}\,, 
\end{split}
\label{eq:sp}
\end{equation}
where $r_\mathrm{rot}$ is the distance from the Galactic center, measured in the frame co-rotating with the spirals arms. The value $H$ is the scale height, $A_\mathrm{sp}$ is the amplitude of the spiral arms, and $R_\Sigma$ is the scale length of the drop-off in density amplitude of the arms. In line with previous simulations \citep{jilkova2012origin, martinez2014radial}, we use only the $n=1$ term, with $C_1=8/3\pi$, and the parameters $K_1, D_1$ and
$\beta_1$ given by:

\begin{eqnarray}
K_1 &=& \frac{m}{r_\mathrm{rot}\sin i},\\
\beta_1 &=& K_1H (1+0.4K_1H), \\
D_1&=& \frac{1+K_1H+0.3(K_1H)^2}{1+0.3K_1H}\,,
\end{eqnarray}
where $m$ and $i$ correspond to the number of spiral arms and pitch angle of the spiral structure
respectively. 

Finally, the term $\gamma$ in Eq. \ref{eq:sp} represents the shape of the spiral structure, which is 
described by the expression:

\begin{equation}
\gamma= m\left[ \varphi -\frac{\ln(r_\mathrm{rot}/r_0)}{\tan i} \right]\,.
\end{equation}
\noindent
The parameter $r_0$ is the reference radius for the adopted mass density of the model ($A_{sp}$), and like \citet{jilkova2012origin}, we choose $5.6$ kpc. The parameters for the spiral arms are listed in Table 1. The five primary parameters that determine the impact of the spiral arms model presented above are: radial scale length, scale height, pitch angle, amplitude of density perturbation (the mass density of the arms) in the Galactic plane at $R_{\odot}$, and angular velocity. The references shown in Table 1 motivate our choices for the parameters of the spiral arm model. 

One of the more important parameters to choose is the number of spiral arms. Some have argued for the Galaxy's spiral pattern to be a superposition of multiple modes with varying densities and different pattern speeds \citep[e.g.,][]{quillen2011structure, lepine2011overlapping, sellwood2014transient}. However, some observations \citep[e.g.,][]{benjamin2005first, churchwell2009spitzer} indicate that the non-axisymmetric component of the old stellar disk is dominated by a two-armed (Scutum-Centaurus and Perseus) spiral pattern outside the bar. Consequently, we choose $m=2$ for the simulations presented in this work. The impact of transient spiral arms will be explored in future work.

\begin{table*}
\begin{center}
\caption{Parameters of the Galactic Potential in this study.\label{table:potential}}
\begin{tabular}{lll}
\tableline \tableline
Property                                           & Value                                               & References \\
\tableline \\
\textbf{Axisymmetric} \\[-0.05cm]
Mass of disk, $M_d$            & $4.5 \times 10^{10}$ M$_{\odot}$                                    & \citet{rix2013milky} \\
Major Axis Scale Length, $a_{disk}$  & $3$ kpc                       & \citet{bovy2015galpy} \\
Minor Axis Scale Length, $b_{disk}$ & $0.28$ kpc                       & \citet{bovy2015galpy} \\
Mass of Bulge, $M_b$   & $0.62 \times 10^{10} \, M_\odot$                     & \citet{jilkova2012origin} \\
Scale Length of Bulge, $b_{bulge}$          &  $0.38$ kpc                                & \citet{jilkova2012origin} \\
Halo Mass, $M_h$                                &  $8 \times 10^{11} M_{\odot}$               & \citet{bovy2015galpy} \\
Halo Scale Length, $a_{halo}$ &  $18$ kpc       & \citet{bovy2015galpy} \\
\textbf{Spiral Arms} \\[-0.05cm]
Spiral Arms Mass Density, $A_{sp}$ & $3.9\times10^7 M_{\odot}$/kpc$^3$ & \citet{jilkova2012origin} \\
Spiral Arms Pattern Speed, $\Omega_{s}$ & $25$ km/s/kpc & \citet{jilkova2012origin} \\
Number of Spiral Arms, $m$ & 2 & \citet{jilkova2012origin} \\
Pitch Angle, $i$ & $15.5 \deg$ &   \citet{jilkova2012origin} \\
Spiral Arms Scale Length, $R_{\Sigma}$ & $2.6 $ kpc &  \citet{jilkova2012origin} \\
Spiral Arms Scale Height, $H$ & $0.18 $ kpc &  \citet{jilkova2012origin} \\
Spiral Arms Initial Orientation w.r.t $\odot$, $\phi_{s,0}$                                & $20\deg    $         &  \citet{jilkova2012origin} \\
Spiral Arms Scale Radius, $r_0$ & $5.6$ kpc &  \citet{jilkova2012origin} \\
\textbf{Bar} \\[-0.05cm]
Bar Mass, $M_{bar}$        & $9.8 \times 10^9 M_{\odot}$                               & \citet{pichardo2004models} \\
Bar Axis Ratios              & $1:0.37:0.256$                           & \citet{pichardo2004models} \\
Bar Pattern Speed, $\Omega_b$           & $43$ km/s/kpc & \citet{bland2016galaxy}  \\
Bar Initial Orientation w.r.t $\odot$, $\phi_{b,0}$                                 & $20\deg    $         & \citet{pichardo2004models} \\
\textbf{GMCs} \\[-0.05cm]
Number of GMCs ($z=0$) & 2500 & \citet{1997ApJ...476..144M, nakanishi2003three} \\

\tableline
\end{tabular}
\end{center}
\end{table*}

\subsubsection{Bar}

The galactic bar is an important component of the potential and, along with spiral arms, has been shown to be an important driver of the substructure in phase space that we see in the Galactic disk today \citep[e.g.,][]{monari2016effects, hunt20184}. Following \citet{pichardo2004models}, we model the bar of the Galaxy with a three-dimensional Ferrer's potential \citep{ferrers1877potentials}, which is represented by the following density: 

\begin{equation}
  \rho_\mathrm{bar}= 
  \begin{cases}
    \rho_0 \left( 1-n^2 \right)^k & n < 1\\
    0 & n \geq 1
  \end{cases}\, ,
  \label{eq:bar}
\end{equation}

\begin{equation}
n^2= x_{rot}^2/a^2 + y_{rot}^2/b^2+ z_{rot}^2/c^2.
\end{equation}
\noindent
The parameters $a$, $b$ and $c$ are the semi-major, semi-minor, and vertical axes of the bar, respectively. The term $\rho_0$ in Eq. \ref{eq:bar}  represents the central density of the bar and $k$ its concentration. We chose $k=1$, since there is an approximation we can use to calculate the forces that does not involve integrals and saves us computational time. We assume that the bar rotates as a rigid body at some pattern speed $\Omega_b$. 

The parameters that describe the bar, such as its pattern speed, mass, orientation, and axes, are still under debate and highly uncertain. The values chosen for this work are listed in Table 1. The total mass in the bulge is estimated to be between $1.4-1.7 \times 10^{10} M_{\odot}$ \citep{bland2016galaxy}. For our models, we assume that the total mass budget in the bulge is $1.6 \times 10^{10} M_{\odot}$. Adopting the values used in \citet{jilkova2012origin}, we assume that the bar mass is $0.98 \times 10^{10} M_{\odot}$ and the rest of the mass budget ($0.62 \times 10^{10} M_{\odot}$) is in the classical spheroidal bulge. We chose to model the bar with a major-axis half-length of $3.14$ kpc, and the three-dimensional axial ratios, based on arguments presented in \citet{pichardo2004models}, are 1:0.37:0.256. The pattern speed of the bar is also uncertain and values in the range $40 - 70 $ km/s/kpc have been quoted in the literature (see \citet{bland2016galaxy} and references therein). For our work, we use the value $\Omega_b = 43$ km s$^{-1}$ kpc$^{-1}$ as presented in \citet{bland2016galaxy}. 
	
\subsubsection{Giant Molecular Clouds}

\citet{spitzer1951possible, spitzer1953possible} proposed that interaction with GMCs could contribute to the vertical heating of the stellar disk. Consequently, for a realistic model of the Milky Way potential, GMCs must be included as they play an important role of heating in the disk. The procedure to model giant molecular clouds in our simulations is similar to that of \citet{gustafsson2016gravitational, aumer2016quiescent}. 

\begin{figure*}
 \includegraphics[width=168mm]{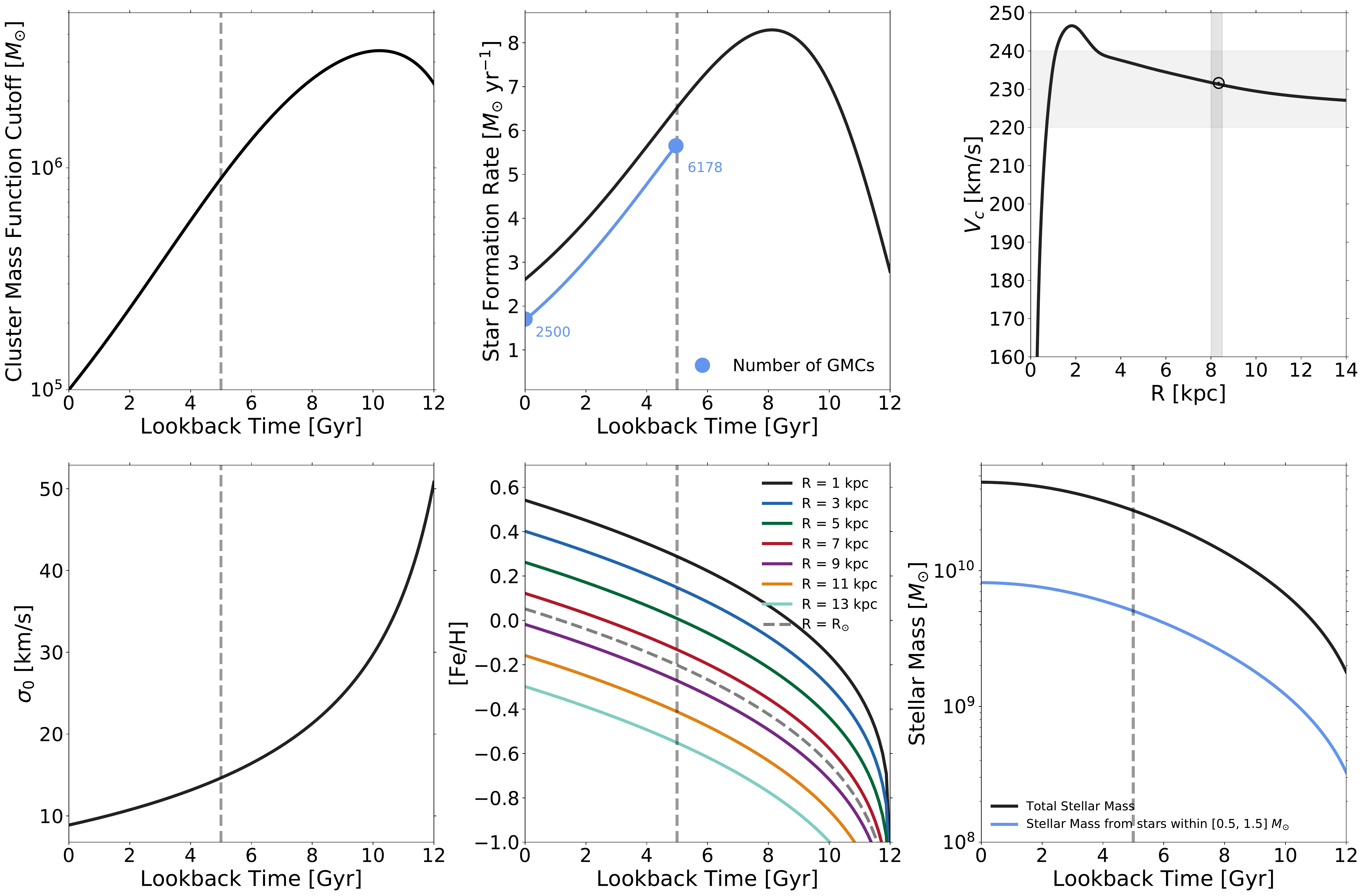}
 \caption{Overview of several model ingredients. Top left: The evolution of the Cluster Mass Function (CMF) as a function of lookback time. The CMF evolution is prescribed using results from \citet{escala2008stability}. \textit{Top center:} The global star formation rate (SFR) evolution of the simulated galaxy derived from the star formation history (SFH) of a Milky Way-sized halo from \citet{behroozi2013average}. The black dashed line shows how long we "actively" dynamically evolve stars in the galaxy -- all stars beyond 5 Gyr are assumed to completely phase-mixed and are instead modelled as a smooth background. The blue line shows the number of GMCs in the simulation, which traces the star formation rate. Top right: The circular velocity curve ($V_c$) that is obtained from the galactic potential model described in Section 2.2 and Table 1 at $t=0$. The horizontal grey box shows the uncertainty of the Milky Way's circular velocity at $R_{\odot}$ and the vertical grey box shows the uncertainty in $R_{\odot}$. Bottom left: The input "birth" velocity dispersion $\sigma_0$ for stars initialized in the galaxy as a function of lookback time. $\sigma_0$ defines the velocity ellipsoid as: $\sigma_0 = \sigma_r = \sigma_z = \sqrt{2} \sigma_{\phi}$. Bottom center: The age-radius-[Fe/H] relation adopted from the work of \citet{frankel2018measuring}. For a given star cluster in the simulation, we assign a value for [Fe/H] based on this relation and derive multidimensional abundances using the method described in Section 2.5. The dashed line shows the evolution of this relation at the solar radius, $R_{\odot}$. Bottom right: The accumulated stellar mass for the SFH as a function of lookback time for all stars (black line) and only stars within [0.5, 1.5] M$_{\odot}$. }
 \label{fig:model}
\end{figure*}

The first tunable parameter in our GMC model is the number of GMCs at $z=0$ in the Galactic disk. \citet{1997ApJ...476..144M} and  \citet{nakanishi2003three} find a total galactic molecular mass of approximately $10^9 M_{\odot}$. The number of GMCs at $z=0$ in the disk is consequently assumed to be 2500 based on the mass function in (Eq. \ref{eq:mf_gmc}). We assume that all the molecular gas mass is distributed in GMCs. In the past, the number of GMCs would likely have been higher due to a higher star formation rate density. For this work, we simply set the number of GMCs as proportional to the SFR at the lookback time divided by the SFR at $z=0$. This simple prescription is plausible since we would expect the $\Sigma_{GMC}$ to trace the $\Sigma_{SFR}$ based on the Kennicutt-Schmidt law \citep{bigiel2008star, kennicutt2012star} for cold molecular clouds, as follows:

\begin{equation}
\Sigma_{GMC} \propto \Sigma_{SFR}^{1/\alpha_{KS}},
\end{equation} where $\alpha_{KS} = 1.0$. 
\noindent
The mass of the GMC is randomly chosen from the following distribution function \citep{hopkins2012structure} with the SFR fraction added to encapsulate the time evolution of the number of GMCs:
\begin{equation}
\label{eq:mf_gmc}
N(M, t) \propto M^{-1.8} dM \times \frac{SFR(t)}{SFR_0},
\end{equation}
where the upper GMC mass limit is $10^7 M_{\odot}$ and the lower mass limit is $10^5 M_{\odot}$ and $SFR_0$ is the star formation rate at $z=0$. The number of GMCs is plotted as a function of time in the top center panel of Figure \ref{fig:model}. 

Now that we have the mass and the number of GMCs as a function of time, we need a prescription for where they will be initialized. The number of GMCs spawned in the galaxy at a given time as a function of galactocentric radius is proportional to the star formation rate as a function of radius at a given time (more on this in Section 2.3). However, we limit this range to $2 $ kpc$< R_{GC} < 10$ kpc due to the dearth of GMCs at larger radii in the case of no truncation. 

In Milky Way-like galaxies, GMCs are preferentially found near spiral arms. To mimic this effect, given a radial annulus and the number of GMCs to be initialized in that annulus, we first find the cylindrical $\phi_{sp,max}$ of the two overdensities in that annulus due to the two spiral arms. The number of GMCs is then evenly split between the two spiral arm overdensities. $\phi_{GMC,birth}$ is determined by drawing from a Gaussian centered on $\phi_{sp,max}$ with a width of $30$ degrees and $R_{GMC,birth}$ is chosen uniformly from the width of the radial annulus. The clouds are initialized in nearly circular orbits around the galaxy with a velocity dispersion ellipsoid of 7 km s$^{-1}$ \citep{larson1979stellar, stark1984kinematics}.

The mass evolution of the GMCs is modeled parabolically with half their lifetime spent growing in mass and the other half losing mass (following \citealt{krumholz2006global, goldbaum2011global}). Each GMC is active in the simulation for roughly a few free-fall times of the cloud, i.e. $40$ Myr, with a mass increasing to a value $M_{GMC,max}$ in $20$ Myr, and then decreasing to zero in another $20$ Myr. The exact form of this parabolic evolution is:

\begin{equation}
M(t) = M_i \, \left(-0.25\left(\frac{t-t_0}{10^7 \, \rm yr}\right)^2 + \left(\frac{t-t_0}{10^7  \, \rm yr}\right)  \right),
\end{equation}

\noindent where $t_0$ is the formation time of the GMC and $t$ is the current time in the simulation. The radius evolution of a GMC given its mass is given by \citet{hopkins2012structure}:

\begin{equation}
R(t) = 20 \, \left(\frac{M(t)}{5\times 10^5 M_{\odot}}\right)^{1/2} \text{pc}.
\end{equation}
Given the mass and the radius, we choose to represent the GMCs as simple Plummer spheres \citep{plummer1911problem} in the simulation. With the full GMC model at hand, at any given time in the simulation, we know how many GMCs there are, where they are, and how large they are. However, calculating the force on a given test particle due to all these GMCs turns out to be computationally prohibitive. Instead, we choose to represent all GMCs present at a given snapshot in the simulation in a kD-tree, allowing us to only focus the force computation on nearby GMCs. Given the position and velocity of a test particle, we sum up the forces due to all the GMCs in a $0.2$ kpc radius sphere around the particle. This recipe leads to order of a percent-level force errors when compared to the actual force by summing over the contribution of every GMC's influence. We hope to improve this in the future by pre-computing the GMC force-field and using cubic spline interpolation to significantly speed up the force calculations. 

\subsubsection{Orbit Integration}
\label{sec:orbit}
We use a fourth-order symplectic integrator presented in \citet{kinoshita1990symplectic} with a fixed time-step $dt=0.5$ Myr to integrate particles in the potential. The reason for choosing this integrator over more popular integrators such as  4th Order Runge-Kutta or a higher order Dormand-Price adaptive time-step integrator is primarily for computational reasons -- Kinoshita's integrator requires fewer force calculations than Runge-Kutta and is more robust than a generic leapfrog integrator. 

The time-step of the integrator will matter most in the case of GMC scattering. Since both GMCs and stars are set on nearly circular orbits, albeit with slightly different velocity dispersions, let us assume that the velocity difference between a random GMC and a star is $\Delta V = 10$ km s$^{-1}$ $\approx 10$ pc/Myr. Since the typical diameter of a GMC is $\approx40$ pc and the fiducial time-step is $0.5$ Myr, we resolve the GMC-star interaction 8 times as the star crosses the GMC ($dt = \frac{1}{8} t_{cross}$). Moreover, we do not expect close encounters such as the one described above to be the primary scattering mechanism in our simulations. As described in \citet{gieles2006star}, in a Milky Way-like galaxy, we expect that the disruption of clusters due to GMC interactions mostly happens over much longer timescales due to repeated interactions with GMCs at longer ranges rather than due to single head-on encounters.

\subsection{Star Formation History}
\label{sec:sfh}
We now describe the model for the star formation of the simulated galaxy. The SFH of the solar neighborhood has been well-studied \citep[e.g.,][]{twarog1980chemical, bertelli2001star, cignoni2006recovering}, with studies suggesting a rather flat SFH over the past 8 Gyr. However, the global SFH of the Milky Way is not as well understood. Instead of recreating the Milky Way's exact SFH, we choose to model the simulated galaxy's SFH based on subhalo abundance matching in cosmological simulations. In particular, we use the following parameterized SFH from \citet{behroozi2013average} assuming a virial mass of $10^{12} M_{\odot}$: 

\begin{equation}
SFR(t) = A \, \left(\frac{t (Gyr)}{C}\right)^B \exp \left(\frac{-t (Gyr)}{C}\right).
\end{equation}

\citet{ting2015prospects} found that the values $A=15.5 M_{\odot} \text{yr}^{-1}$, $B=2$, and $C=2.7$ Gyr fit the evolution of the stellar mass surface density at the Sun's location best, and therefore those are the values adopted in this work. The normalization is adjusted to make sure that the total stellar mass agrees with $M_{z=0} = 4.5 \times 10^{10} M_{\odot}$. For the purposes of the work presented in this paper,  we only evolve stars born in the mass range $[0.5, 1.5] M_{\odot}$, which account for roughly $18 \%$ of the stars in a stellar population if we assume a Kroupa IMF with a low mass limit of $0.08 M_{\odot}$ and a high mass limit of $100 M_{\odot}$. The mass range $[0.5, 1.5]$ $M_{\odot}$ encompasses the bulk of the stars that we are interested in.  More massive stars are both rare and short-lived, while less massive stars are too faint to have high quality {\it Gaia} parameters over a significant volume. 

The radial growth of the Milky Way is specified using the relation found observationally by \citet{van2013assembly} while studying nearby Milky Way-like galaxies: 

\begin{equation}
\label{eq:growth}
R_{\star} \propto M_{\star}^{0.27}.
\end{equation}
\noindent
Unfortunately, the scale length of the star formation rate is a significantly harder quantity to constrain for the Milky Way. Most of our constraints are for the stellar disk scale length but not the SFR scale length. Consequently, like T15, the SFR scale length is adopted from the extragalactic study of NGC 6946 \citep{schruba2011molecular}, a Milky Way-like galaxy, and is assumed to be $2.6$ kpc at $z=0$. The evolution of this scale length is assumed to trace the evolution of the effective radius of the Milky Way, which is proportional to $M_{\star}^{0.27}$ (Eq. \ref{eq:growth}). T15 showed that the form showed in Eq. \ref{eq:growth} coupled with an analytic model for radial migration \citep{freeman2002new} leads to a reasonable value of the scale length of the stellar disk indicated that the chosen SFR scale length is a robust choice. 

\subsection{Star Cluster Model}
\label{sec:sc}
In this section we describe our model for how individual stars are initialized into clusters, a subgrid model for mimicking intra-cluster \textit{N}-body interactions, and a simple recipe for star cluster disruption. The fundamental assumption of our star cluster model is that all stars born in the disk are born in bound or unbound star clusters \citep{lada2003embedded}. We discuss this assumption in detail in Section \ref{sec:discussion} and leave the exploration of alternative clustered star formation recipes for future work. 

\subsubsection{Mimicking \textit{N}-body Dynamics with Test Particles}

The computational cost to perform direct \textit{N}-body simulation of the entire Galactic disk is not currently feasible. The largest such simulations are of the order $10^6$ M$_{\odot}$ \citep[e.g.,][]{wang2016dragon} but that is nowhere near the dynamic range needed to simulate the whole Galaxy on a star-by-star basis. Moreover, the transition between gas-rich and gas-free clusters is poorly understood \citep{krumholz2018star}, and hence choosing realistic initial conditions is challenging and ultimately requires {\it ab initio} hydrodynamic simulations, driving the computational cost even higher. We have therefore decided to simulate the dynamical evolution of the galaxy with orbit integration of test particles coupled to a method of initializing the clusters in a way that roughly approximates their evolution in the disk. 

\par In order to set the initial conditions of the star particles, we ran a series of \textit{N}-body star cluster simulations in the galaxy potential described in Section \ref{sec:galaxy_pot} using the Astrophysical MUltipurpose Software Environment (AMUSE) framework \citep{portegies2011amuse}. AMUSE is meant to couple different astrophysical simulation codes to evolve complex systems involving physical processes of very different scales \citep{whitehead2013simulating}. For this work, we combine an \textit{N}-body code (NBODY6; \citet{aarseth1999nbody1}) with a model for the galactic potential that was described above in Section \ref{sec:galaxy_pot}. We use the ROTATING BRIDGE \citep{fujii2008evolution, martinez2014radial} package to interface the intracluster \textit{N}-body forces and the tidal field of a galaxy with rotating, non-axisymmetric components. The reader is referred to Appendix A of \citet{martinez2014radial} for a thorough explanation of ROTATING BRIDGE.

\begin{figure*}
 \includegraphics[width=168mm]{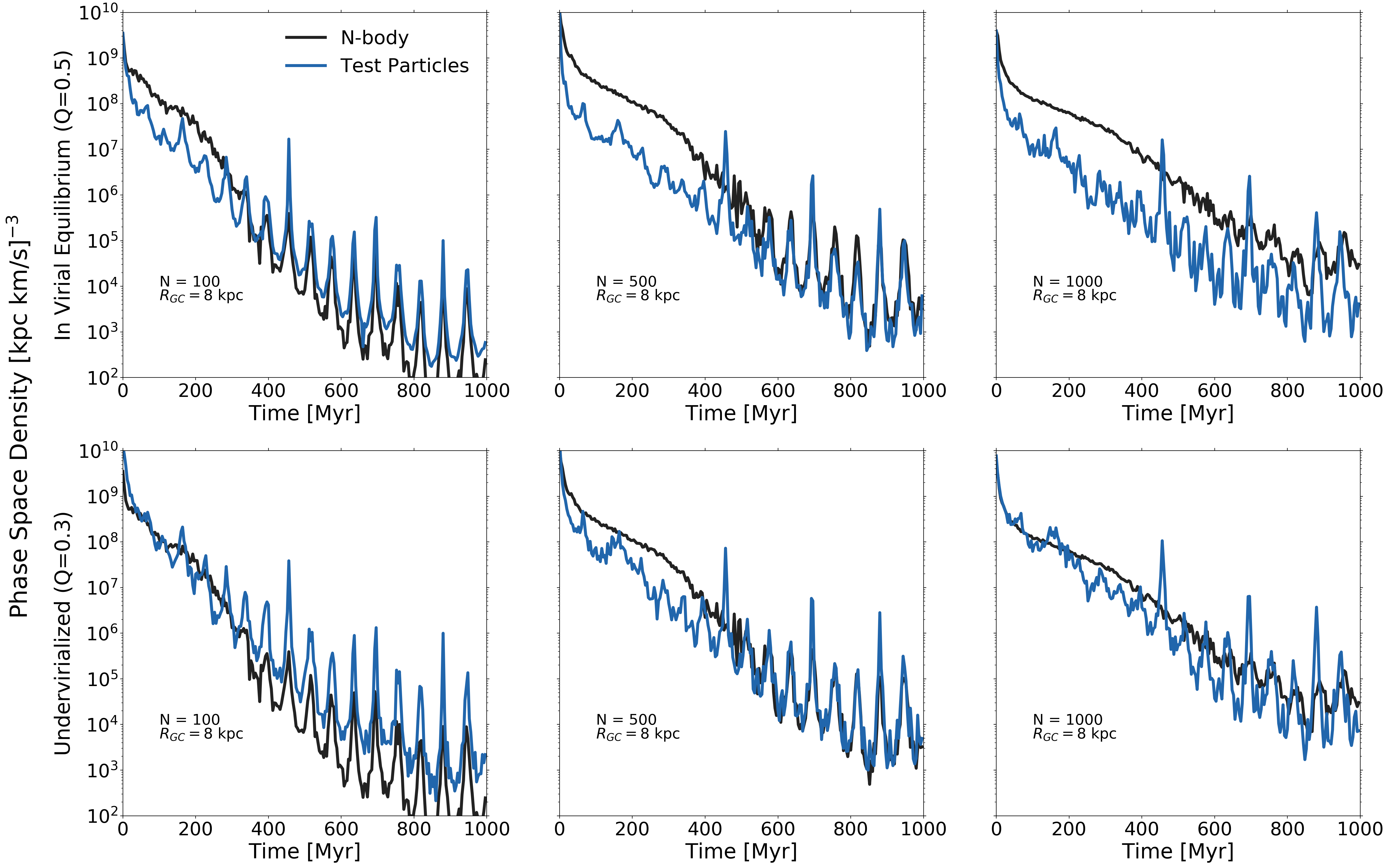}
 \caption{Mean phase space density as a function of time comparing direct \textit{N}-body (black) to test particle integration (blue) for different star clusters of different masses. The phase space densities are computed using EnBiD \citep{sharma2006multidimensional}. The top panels show the test-particle clusters initialized in virial equilibirium ($Q=0.5$) and the bottom panels show  test-particle clusters initialized slightly undervirialized ($Q=0.3$). The N-body calculations are the same in the top and bottom panels.  The top panels show a significant disagreement in the evolution of the phase space densities at early times -- star clusters modeled self-consistently are more bound whereas the test particles disperse quickly.  In contrast, the undervirialized clusters in the bottom panels show much better agreement with the direct N-body calculations.}
 \label{fig:nbody_test_comp}
\end{figure*}

We use the phase space density as our primary metric to compare the evolution of star clusters between a full \textit{N}-body calculation and a test particle calculation. The numerical calculation of phase space densities is described in Section \ref{sec:pd}. Figure \ref{fig:nbody_test_comp} shows the phase space density evolution for clusters of different masses for the fully self-consistent \textit{N}-body case and the test particles case. The figure shows the mean of the phase space densities of 10\% of particles with the highest phase space densities since we want to compare the most bound parts of the two clusters (we find no qualitative differences when we use the median of the phase space densities of all particles as opposed to the top 10\% mean). As we can see in the top panels, due to self gravity, the \textit{N}-body cluster stays bound for longer than the test particles cluster. In order to mimic the effects of self-gravity, we undervirialize the star clusters in the test particles case. The virial ratio, $Q$, is defined as: $Q = \frac{T}{|W|}$, where T is the kinetic energy of the ensemble and W is the potential energy ($Q=0.5$ implies virial equilibrium). After numerous experiments, we found that setting $Q=0.3$ led to behavior that more closely resembled the \textit{N}-body calculations (shown in the bottom panel).  

The approach presented here carries obvious limitations. However, there are a variety of potential enhancements to the simple model presented here that are subject of future work. One could, for instance, set the undervirialization to be a function of the mass of the star cluster and its birth galactocentric radius. A thorough exploration of these recipes is left for future work. It is also worth emphasizing that this test is carried out in isolation with only stars and no gas. However, star clusters are born in gas clouds and, as discussed in the Introduction, the transition from gas-rich to gas-free population of clusters is still uncertain. As we will see later in this section, the virializations (and, consequently, the velocity dispersions of clusters) will need to be further modified to realistically mimic star cluster disruption in the Milky Way.

\subsubsection{Cluster Mass Function}

The cluster mass function (CMF) plays a critical role in determining the clumpiness of phase space and chemical space in the Milky Way disk. Following \citet{lada2003embedded, ting2015prospects}, we characterize the CMF with the power law: 

\begin{equation}
\frac{dN}{dM} \propto M^{-\alpha},
\end{equation}
where $\alpha = 2$. The lower mass cutoff for the fiducial model is fixed to be $50 M_{\odot}$. The higher mass cutoff has a dramatic impact on the clumpiness in both the phase space and chemical space of the Milky Way. The largest open clusters observed in the Milky Way today are Westerlund  1 \citep{brandner2008intermediate}, Berkeley 39 \citep{bragaglia2012searching}, and Arches \citep{espinoza2009massive}, all with a mass few times of $10^4$ M$_{\odot}$. Consequently, following T15, we conservatively choose the high mass cutoff at $z=0$ to be $10^5 M_{\odot}$ (similar also to \citet{bland2010long}'s choice of $2 \times 10^5 M_{\odot}$) to compensate for rapid mass loss at very early ages. 

The evolution of the CMF over time is less certain. \citet{escala2008stability} present a model to approximate the maximum cluster mass by studying the gravitational instabilities in disks. They found that we can estimate the evolution of the maximum cluster mass using the relation 

\begin{equation}
M_{max, cluster}(t) \propto \eta ^2 M_{gas}(t),
\end{equation}
where $\eta = \frac{M_{gas}}{M_{gas} + M_{\star}}$ and $M_{gas}(t)$ is the evolution of the gas mass of the Milky Way. We assume $M_{dynamical} = M_{*}$ because we assume the Milky Way is mostly "maximal", i.e., the potential is dominated by stellar mass in the disk. Assuming the gas surface density at the Sun's location to be $\Sigma(R_0, z=0) = 13 M_{\odot}/\text{pc}^2$ \citep{flynn2006mass} and the scale length $R_{gas} = 4.2$ kpc \citep{schruba2011molecular}, the total gas mass in the Milky Way at $z=0$ is $9.7 \times 10^9 M_{\odot}$. Given this, we can calculate the evolution of the gas mass $M_{gas}(z)$ using the Kennicutt-Schmidt relation with $\alpha_{KS} = 1.5$ and the SFR evolution described in the previous section. The value for $\alpha_{KS}$ is higher here compared to the value presented in Section 2.2.4 for GMCs since the latter is for cold molecular clouds and the former is the disk-averaged star formation law \citep{kennicutt2012star}. The normalization for this relation is set by fixing the maximum cluster mass to $10^5 M_{\odot}$ at $z=0$ as discussed above. The evolution of the CMF high mass cutoff is shown in the upper left panel of Figure \ref{fig:model}. 

Given the SFR as a function of radius and time and the cluster mass function as a function of time, the galaxy is split up into annuli of $0.1$ kpc and the number of stars formed in each annuli based on the SFH is divided by the average mass of stars in a cluster (around $500 M_{\odot}$ for a CMF cutoff of $10^6$ M$_{\odot}$). This quantity gives us the approximate number of star clusters to be initialized. The clusters in each annulus are then drawn from the CMF shown above. 

\subsubsection{Cluster Structure \& Early Evolution}

As discussed in the Introduction, the initial conditions of star clusters in the disk is highly uncertain and hotly debated due to the uncertainties in the transition between the gas-dominated and the gas-free phase and the scale of the structure of the ISM \citep[e.g.,][]{lada2003embedded, goodwin2006gas, 2008MNRAS.384.1231B, pfalzner2009universality, bressert2010spatial, assmann2011popping, parker2012characterizing, kuhn2014spatial, pfalzner2014evolutionary, longmore2014formation, elmegreen2018dispersal}. To recap, the classical view is that most star formation happens in clustered environments ("embedded clusters") and these evolve into classical open clusters after gas expulsion. \citet{lada2003embedded} found that only a small fraction survive the embedded phase (the  "infant mortality" scenario), where around 90\% of all star clusters are disrupted on very short timescales ($<10$ Myr). However, recent work \citep[e.g.,][]{bressert2010spatial, kruijssen2012fraction, elmegreen2018dispersal} argues for a hierarchical, spatially correlated ISM. The structure of the ISM in this scenario implies that instead of $90\%$ of star clusters being disrupted instantly after being born, $90\%$ of the stellar mass is in the unbound part of the hierarchical structure and escapes soon after gas dispersal. For the rest of this work, we assume the classical \citet{lada2003embedded} view, where all stars are born in either bound or unbound clusters and the existence of a lower mass cutoff for the CMF. In future work, we plan to model the spatial structure of a hierarchical ISM to compare these two scenarios in detail.

To initialize a star cluster in the galaxy, we require three parameters: the mass of the cluster, the radius of the cluster, and the virial ratio of the cluster, $Q$, which sets the internal velocity dispersion. The CMF and the SFR provide us with the masses of the star clusters born in the galaxy. However, determining the initial radii of star clusters is rather complicated. There are broadly three different kinds of young star clusters that have been observed in the Milky Way: (1) embedded clusters, (2) associations (leaky clusters in \citet{pfalzner2009universality}), and (3) young massive clusters ($<10$ Myr and >1000 $M_{\odot}$ pc$^{-3}$).

The formation channels of these different types of clusters is still debated and the distinctness of these different types is contested \citep[e.g.,][]{pfalzner2009universality}. However, since these different clusters have a different imprint on the mass-radius relation of clusters in the Milky Way, we choose to model them separately for the sake of simplicity. Broadly, embedded clusters evolve into classical open clusters if they survive gas expulsion. If they do not survive gas expulsion, they are unbound from very young ages and dissolve in the galaxy much quicker. Young Massive Clusters (YMCs; referred to as starburst clusters in \citet{pfalzner2014evolutionary}) are very high mass, compact clusters. YMCs are common in nearby starburst galaxies but only a handful are observed in the Milky Way today. Observational studies suggest that YMCs are bound \citep[e.g.,][]{longmore2014formation}. Finally, associations (leaky clusters in \citet{pfalzner2014evolutionary}) have a mass similar to YMCs but are orders of magnitude less dense. Most associations are thought to be unbound. \citet{gieles2011distinction} argue that the distinction between an open cluster and an association can be determined by the ratio:  $t_{age}/t_{dyn}$ -- if the ages of the stars are greater than the dynamical time of the system, the stars must be bound. 

\begin{figure}
 \includegraphics[width=84mm]{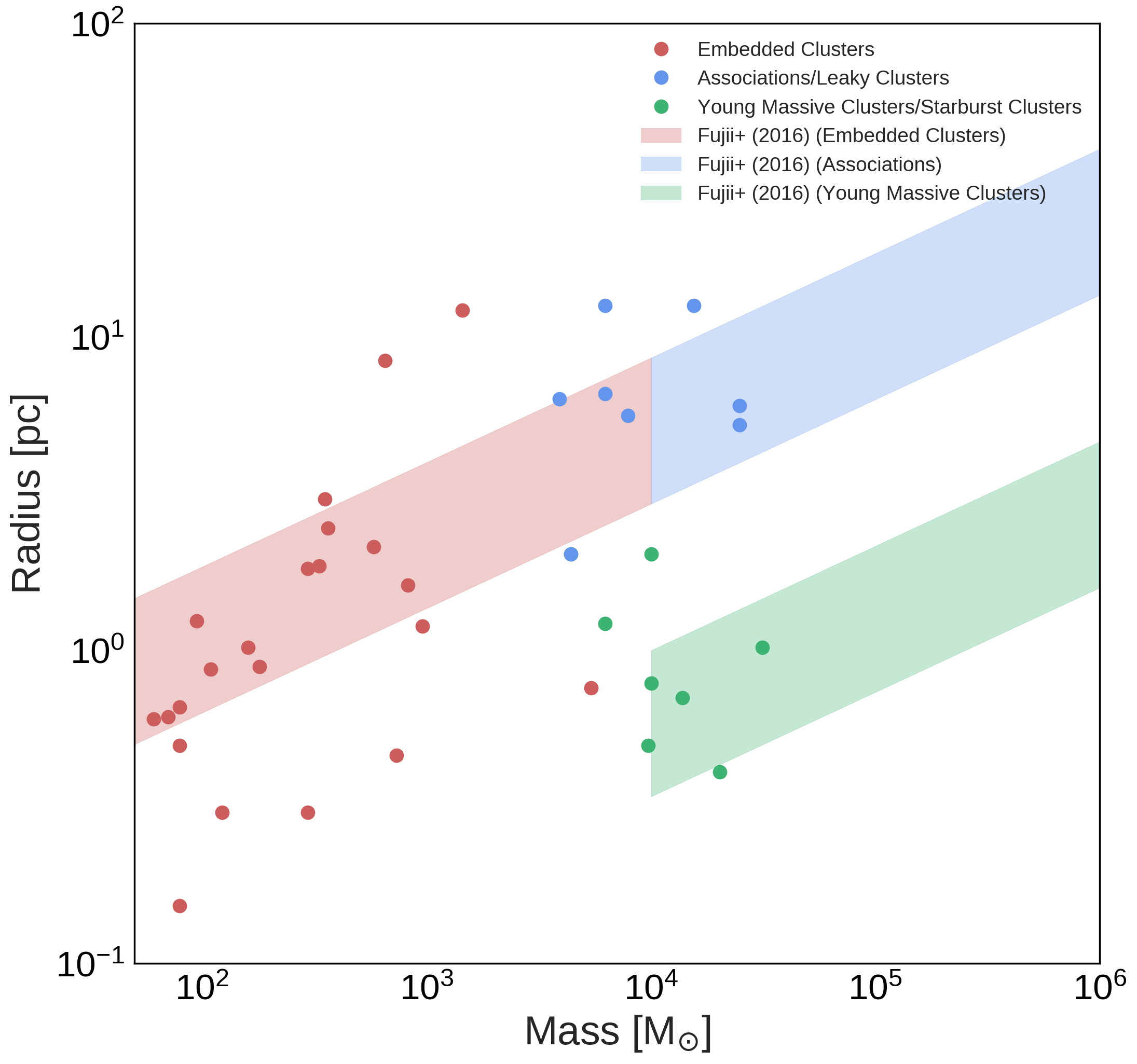}
 \caption{Initial mass-radius relation based on Eqs. 19 and 20 for embedded clusters, associations, and young massive clusters with corresponding observations. The red and blue shaded regions shows Eq. \ref{eq:mr_oc} from $t=1$ Myr  to $t=5$ Myr for masses $< 10^4$ and $> 10^4$ M$_{\odot}$ respectively, the green shaded region shows Eq. \ref{eq:mr_ymc} from $t=1$ Myr  to $t=5$ Myr. Overplotted are embedded cluster observations in red, unbound associations in blue, and young massive clusters in green. The ages of all clusters presented here are between 1 and 5 Myr.}
 \label{fig:mass_radius}
\end{figure}

Figure \ref{fig:mass_radius} shows the observed mass-radius relation compiled in \citet{fujii2016formation} for clusters with ages less than 5 Myr. The red dots show classical embedded clusters, the blue dots show associations, and the green dots show YMCs. There is a lot of scatter in this parameter space for two primary reasons: (1) different regions in this space are sensitive to different feedback mechanisms \citep[reviewed in][]{krumholz2018star}, and (2) the observational difficulty associated with defining an embedded cluster while it is still in its natal cloud. Consequently, for the purposes of this work, we use the simple parameterized mass-radius relations presented in \citet{fujii2016formation} (FPZ hereafter). 

FPZ argue that associations ($M > 10^4 M_{\odot}$) and smaller embedded clusters ($M < 10^4 M_{\odot}$) lie on the same mass-radius track and YMCs ($M > 10^4 M_{\odot}$) lie on a different mass-radius track because YMCs are much denser. The mass-radius relation for associations and embedded clusters derived in FPZ is given as:
%

\begin{equation}
\label{eq:mr_oc}
r_h = \left(\frac{t_{dyn}}{2\times 10^4 \rm yr}\right) \left(\frac{M}{10^6 M_{\odot}}\right)^{1/3} \text{pc},
\end{equation}
\noindent
where $M$ is the mass, $r_h$ is the half mass radius, and $t_{dyn}$ is the dynamical time. Further details on how this expression is derived can be found in FPZ. 

For young massive clusters, the mass-radius relation is more involved to derive due to the different physics at play before core collapse and after core collapse. After making some simplifying assumptions and using results from \citet{2011MNRAS.413.2509G}, FPZ arrive at the approximate the mass-radius-age trend for young massive clusters: 

\begin{equation}
\label{eq:mr_ymc}
r_h = 0.0158 \left(\frac{M}{M_{\odot}}\right) ^{1/3} \left(\frac{t_{age}}{\text{Myr}}\right)^{2/3} \text{pc}.
\end{equation}
\noindent
Further details on how this expression is derived can be found in FPZ. Equation \ref{eq:mr_oc} \& \ref{eq:mr_ymc} are both plotted in Figure \ref{fig:mass_radius} from $1 < t < 5$ Myr (shown as the shaded regions in different colors for the different types of clusters). The relations are in broad agreement with observations. 

The above describes how one could pick an initial radius for the three different types of clusters given a mass. However, to initialize the cluster in phase space, we need a value for the virialization parameter. In Section 2.4.1, we demonstrated that slightly undervirializing test particles in a cluster broadly reproduces the phase space evolution of clusters with different masses. However, from \citet{lada2003embedded}, we expect that not all of the initialized clusters will end up bound ("infant mortality") due to gas expulsion. We need a metric that can dictate what proportion of star clusters end up bound and what proportion end up unbound. 

First we consider clusters with masses greater than $10^4$ M$_{\odot}$. Since YMCs require such a high surface density, they are only really found in the inner parts and the spiral arms of a Milky Way-like galaxy. There is strong evidence for YMCs being bound systems from very young ages \citep{longmore2014formation}. In contrast, \citet{portegies2010young,gieles2011distinction} argue that associations are unbound from birth. 
 
Consequently, we use the cluster formation efficiency (CFE) as our primary constraint on the proportion of star clusters above $10^4$ M$_{\odot}$ that are bound (are initialized on the YMCs track) and that are unbound (are initialized on the associations track). The CFE is just the fraction of star formation that occurs in bound star clusters \citep[e.g.,][]{bastian2008star, goddard2010fraction, kruijssen2012fraction}. For $M > 10^4$ M$_{\odot}$, we assume a fidcuial value of 10\% for the CFE 
-- $10\%$ of star clusters above that mass will be initialized on the YMC sequence and the other $90\%$ will be initialized on the associations sequence.

\begin{figure}
 \includegraphics[width=84mm]{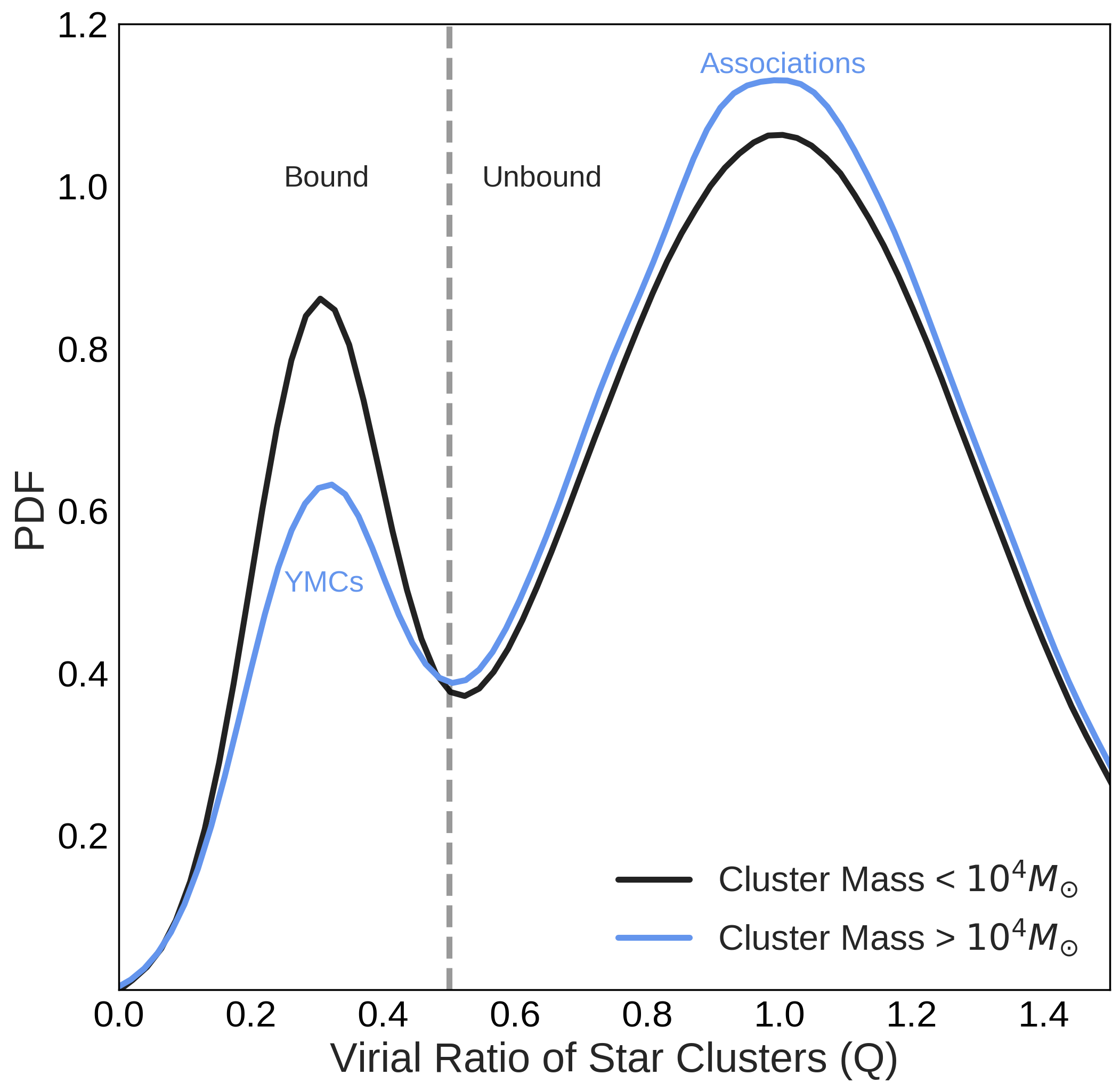}
 \caption{The distribution of the virialization parameters, $Q$, for the star clusters initialized in the simulated galaxy. The tuning of the virializations and the corresponding effective star formation efficiencies is driven by calculations of the cluster formation efficiency in Milky Way-like galaxies \citep{kruijssen2012fraction}.}
  \label{fig:virial_sfe}
\end{figure}

Since we expect some variation in how YMCs evolve, we initialize bound YMCs with virializations drawn from the gaussian $\mathcal{N}(Q=0.3, \sigma_Q = 0.1)$. To initialize unbound clusters due to gas expulsion, we choose a broad spread for the overvirializations with the gaussian $\mathcal{N}(Q=1.0, \sigma_Q = 0.3)$ using arguments similar to those used in previous works \citep{baumgardt2007comprehensive}. Early \textit{N}-body simulations overvirialized star clusters based on an effective star formation efficiency given by $\epsilon = \frac{M_{*}}{M_{gas}+ M_{*}}$. After gas expulsion, making the crude assumption that $M_{gas}$ is completely removed on a very short timescale, we can clearly see that the new kinetic energy $T = \epsilon T'$ and the potential energy $W' = \epsilon^2 W$. Therefore, the virial ratio can be chosen based on the relation $Q \propto \frac{1}{\epsilon}$. The effective star formation efficiency is highly uncertain due to different feedback mechanisms. The chosen virialization distributions roughly correspond to a mean effective star formation efficiency of 0.3 with a large scatter that accounts for the different strengths of feedback processes. 

For clusters with masses less than $10^4 M_{\odot}$, there is not a clear distinction between the bound/unbound sequence in the mass-radius plane like for the high mass clusters. FPZ argue the radius as a function of mass and age for embedded clusters ($M < 10^4$ M$_{\odot}$) is the same as for associations  (Eq.\ref{eq:mr_oc}). Figure \ref{fig:mass_radius} shows this mass-radius relation. However, if we assume that the mass-radius relation for embedded clusters is the same as for associations, which were all initialized as unbound, we need to increase the bound fraction for the smaller embedded clusters to account for this discrepancy, since we know from observations that embedded clusters do end up evolving into bound systems (open clusters). 

Consequently, we choose a 20\%/80\% split between bound and unbound clusters for these smaller clusters. More specifically, we initialize star clusters with $M_{\odot} < 10^4$ M$_{\odot}$ with a virialization $\mathcal{N}(Q=0.3, \sigma_Q = 0.1)$ $20\%$ of the time (a lot of these systems will evolve into classical open clusters) and with $\mathcal{N}(Q=1.0, \sigma_Q = 0.3)$ $80\%$ of the time (a lot of these systems will get phase-mixed on very short timescales). Another reason for a slightly higher fraction of bound systems compared to more massive systems is that YMCs are exceedingly rare in Milky Way-like galaxies \citep{portegies2010young} (especially in the last 5 Gyr, the timescale for which we dynamically evolve stars) implying that a higher bound fraction for smaller clusters is warranted. The impact of parameterizing the bound fraction as a function of environment that the star cluster was born in and how long ago it was born will be explored in future work. The different virial ratio distributions for the clusters in the two different mass bins are shown in Figure \ref{fig:virial_sfe}. The different weighting of the gaussian distributions leads to about $8 \%$ of star clusters with masses greater than $10^4 M_{\odot}$ and about $11\%$ of star clusters with masses less than $10^4 M_{\odot}$ to have virial ratios less than $Q=0.3$. 

Now that we have a mass, radius, and virialization, we can initialize all the stars belonging to a cluster into a Plummer sphere.

\subsubsection{Cluster Orbit Initialization}
\label{sec:cluster_init}

The stars belonging to a cluster are initialized with a Kroupa IMF \citep{kroupa2001variation} from $0.08$ M$_{\odot}$ to $100$ M$_{\odot}$. They are then placed in a nearly circular orbit around the galaxy. The birth velocity dispersion as a function of time in the simulation to be added to all the stars belonging to a cluster, similar to \citet{aumer2017migration} is given by: 

\begin{equation}
\sigma_{CL}(t) = \sigma_0 \left[\frac{t+t_1}{2.7 \text{Gyr}}\right]^{-0.47} - 15 \,\text{km s}^{-1}, 
\label{eq:init_dv}
\end{equation}

\noindent where $\sigma_0 = 51$ km s$^{-1}$, $t_1 = 1.57$ Gyr. The input velocity dispersion is shown in the top left panel of Figure \ref{fig:obs} as the grey line. $\sigma_{CL}(z=0) \approx 8$ km s$^{-1}$. The specific velocity dispersions are given by $\sigma_r = \sigma_0 = \sigma_z = \sqrt{2} \sigma_{\phi}$. As discussed in \citet{aumer2017migration}, the motivation for this age-velocity dispersion relation comes from \citet{wisnioski2015kmos3d}, who studied the observed H$\alpha$ dispersions and found $\sigma_{H\alpha} \propto (1+z)$ for disk galaxies in the KMOS3D survey. The power law came from an approximation to this proportionality under the crude assumption that the kinematics of young stars follow H$\alpha$ kinematics. 

\subsection{Chemistry}
\label{sec:chemistry}

An important aspect of the model is having realistic multidimensional chemical abundances for the stars in the simulation. In principle, one could model chemical evolution to predict the chemical abundances of stars \citep[e.g.,][]{kobayashi2006galactic}, but due to uncertainties in models and since these models only provide tracks but do not allow us to sample from the full multidimensional abundance space, we choose a semi-empirical approach instead. 

We fit a Gaussian Mixture Model (GMM) with 40 components to all of the APOGEE stars with abundances derived using the Payne \citep{ting2018payne}. Formally, a GMM is a linear mixture of individual multivariate Gaussians defined as: 

\begin{equation}
p(x) = \sum_{k=1}^K \pi_k \mathcal{N}(x | \mu_k, \Sigma_k),
\end{equation}
\begin{equation}
\mathcal{N}(x | \mu, \Sigma) = \frac{1}{(2\pi)^{D/2}} \frac{1}{|\Sigma|^{1/2}} \exp \left(-\frac{1}{2}(x-\mu)^T \Sigma^{-1} (x-\mu)\right),
\end{equation}

where $\pi_k$ are the mixing coefficients which all add up to 1 and $K$ is the number of multivariate Gaussians in the GMM. The optimal values for the means and covariances of each GMM component are found using the expectation maximization implemented in SciKit-Learn \citep{pedregosa2011scikit}.

The procedure for deriving the chemical abundances is as follows: an empirical relation for [Fe/H]-age-radius is adopted from \citet{frankel2018measuring} shown in Figure \ref{fig:model}, which is based on fitting the APOGEE red clump sample. Using the fitted GMM and given a star cluster's birth time and birth radius, one can then determine the [Fe/H] from the Frankel relation. To get abundances for other elements, we sample from the conditional distribution given a value for [Fe/H]; since the GMMs are comprised of multivariate normals (MVNs), this conditional distribution also turns out to be an MVN, making the computation straightforward. We can write the conditional distribution for each component $k$ as (where $x_B = $[Fe/H]):

\begin{eqnarray}
p_k(\mathbf{x}_A | \mathbf{x}_B) =
\frac{p_k(\mathbf{x}_A, \mathbf{x}_B)}{p_k(\mathbf{x}_B)} =
\mathcal{N}(\mathbf{x_A}|\boldsymbol{\mu}_{kA|B}, {\mathbf{\Lambda}_{kAA}}^{-1}), \\
\boldsymbol{\mu}_{kA|B} = \boldsymbol{\mu}_{kA} - \mathbf{\Lambda}_{kAA}^{-1}  \mathbf{\Lambda}_{kAB}  (\mathbf{x}_B - \boldsymbol{\mu}_{kB}).
\end{eqnarray}
Consequently, the total conditional distribution can be written as:
\begin{eqnarray}
p(\mathbf{x}_A | \mathbf{x}_B) = \sum_{k=1}^{K} \pi'_k p_k(\mathbf{x}_A | \mathbf{x}_B), \\
\pi'_k =
\frac{\pi_k \mathcal{N}(\mathbf{x}_B|\boldsymbol{\mu}_{kB},\mathbf{\Sigma}_{kBB})}
{\sum_k \mathcal{N}(\mathbf{x}_B|\boldsymbol{\mu}_{kB},\mathbf{\Sigma}_{kBB})}.
\end{eqnarray} 

Using the technique described above, each star cluster is given its unique chemical tag. The fiducial uncertainty in the abundances is assumed to be 0.03 dex but we vary this value to study the impact of precision on the recoverability of stars born in the same cluster. Moreover, individual abundance uncertainties are assumed \textit{not} to be correlated. However, as shown in \citet{ting2015apogee}, due to the large dimensionality of the chemical space, if the covariances are not taken into account, the effective uncertainty could be much smaller. Lastly, the chemical model presented above does not take into account the spatiotemporal correlation between abundances of star clusters as discussed in detail in \citet{krumholz2017metallicity}. These model enhancements will be the subject of future work. 

\subsection{Background Population}
\label{sec:bg}

The model now has most of the pieces assembled: a realistic potential, SFH, chemical model, and a star cluster formation and disruption model. Due to computational constraints and the breakdown of our assumptions at higher redshift (for e.g., no mergers), we choose to run the simulation for the last 5 Gyr in the disk. The number of stars in the mass range $[0.5, 1.5]$ M$_{\odot}$ is approximately $4 \times 10^9$. Running the simulation for this long takes approximately 400,000 CPU hours with most of the time-consuming code optimized with Cython. 

We add older stars (5 to 12 Gyr) by making the assumption that they are completely phase mixed. Consequently, instead of dynamically evolving these stars, we simply add them as a "smooth" background. The same procedure of spawning star clusters based on the SFR in the disk, drawing from the cluster mass function, and placing them in the galaxy described above is used to create the background. The only difference here is that the stars born in a single cluster are not placed in that singular location in the galaxy. Accounting for radial migration and the eventual phase-mixing, the stars in a given cluster spawned for the background are randomly placed between an annulus determined by its birth radius and the average radial migration given by the following formula from \citet{frankel2018measuring}: 

\begin{equation}
<R - R_{birth}> = 3.6 \text{ kpc } \left(\frac{t}{8 \text{Gyr}}\right)^{1/2}.
\end{equation}
The azimuthal angle for the stars are randomly assigned. The thin-disk/thick-disk transition is heavily dependent on the merger history of the Milky Way. Since the merger history of the Milky Way is actively under debate, we assume that the scale height of the galaxy as a function of age roughly increases proportional to the age from 200 to 900 pc \citep{bovy2012milky}. However, we know from simulations \citep{martig2014dissecting, minchev2016relationship} that the disk is flared at old ages ages and large radii. The approximation that the scale height varies smoothly at the solar radii and lower radii though is a reasonable one. Chemistry is assigned using the method in Section \ref{sec:chemistry} -- using the [Fe/H] - radius - age relation and drawing from the GMM. 

Choosing velocities for the background stars is not straightforward since the form for the initial velocity dispersion shown in Eq. \ref{eq:init_dv} is for stars that have just been born and not stars that have already been dynamically heated. Consequently, we adopt velocity dispersions as a function of age for the background from \citet{holmberg2009geneva} since the background is a static population and not dynamically evolved. 

\begin{figure*}
 \includegraphics[width=168mm]{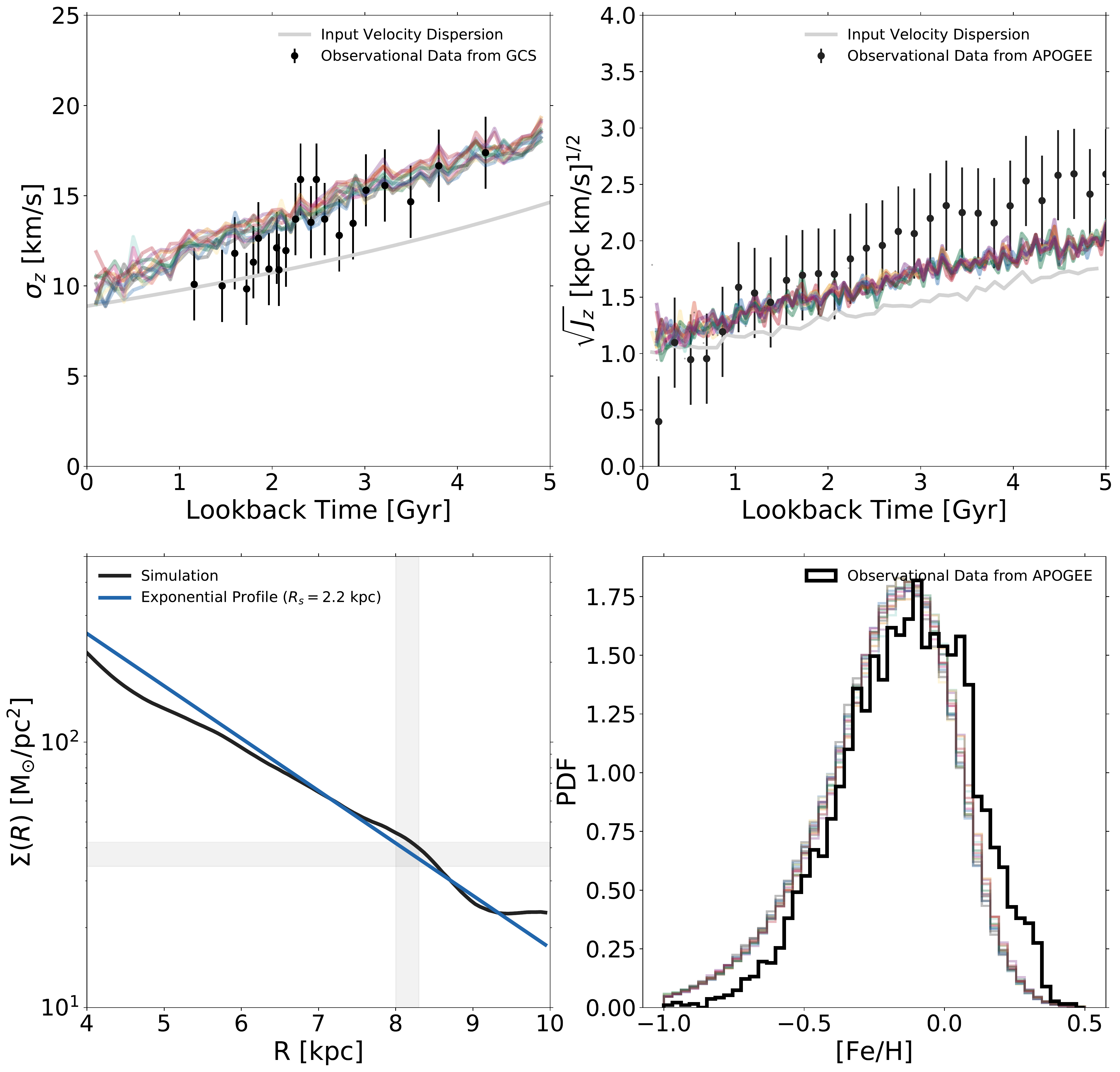}
 \caption{Observational tests comparing the global properties of the simulated galaxy to the Milky Way. Top left: The age-$\sigma_z$ relation seen in the Milky Way (black dots) and various solar neighborhoods in the simulated galaxy (the lines with different colors and a lower transparency). The age-$\sigma_z$ relation is a fundamental constraint on the vertical heating history of the Milky Way. The agreement here implies that the secular evolution of the disk mass and various scatterers, most importantly the GMC model, are broadly consistent with observations of the Milky Way. Top right: The age-$J_z$ relation seen in the Milky Way (black dots), various solar neighborhoods in the simulated galaxy (the lines with different colors and a lower transparency), and the age-birth $J_z$ relation of stars in the simulation. The age-$J_z$ relation is a robust measure of the vertical scattering in the Milky Way. The results shown here imply that the vertical scattering in the simulations due to GMCs is slightly weaker than expected from observations. Potential reasons for this discrepancy and future directions that will explore this discrepancy in more detail are discussed in Section 3. Bottom left: The stellar surface density profile comparison between the simulations and observations. The black line shows an exponential profile with a scale length of $2.2$ kpc and normalized with a solar value of $\Sigma_{R_{\odot}} = 38$ M$_{\odot}$/pc$^2$ and the blue line shows the simulation. The simulated galaxy's surface density profile globally agrees with the Milky Way's density profile implying that our model for the star formation history and the galactic potential are consistent with that of the Milky Way. Bottom right: The distribution of [Fe/H] in the solar neighborhood. The black line shows data from the APOGEE survey \citep{2017AJ....154...94M} and the different lines show the [Fe/H] distribution in the same volume in the different solar neighborhoods of the simulation. The simulated galaxy solar neighborhoods are slightly more metal-poor than observations. The reasons for this discrepancy are discussed in Section 3.1.}
  \label{fig:obs}
\end{figure*}

\subsection{A Mock Catalog of the Solar Neighborhood}
\label{sc:phot}

In order to compare the simulations to observations, we must take into account various selection effects and observational uncertainties. Here we focus on reproducing a \textit{Gaia} DR2-like solar neighborhood observation. 

To mimic observational selection effects it is necessary to model the photometry of the simulated stars. Given masses of the stars in each cluster from Section \ref{sec:cluster_init} and the age and [Fe/H] from Section \ref{sec:chemistry}, we use the MIST stellar evolutionary tracks \citep{choi2016mesa} to predict the stellar surface temperature, surface gravity, and luminosity. We then use bolometric corrections based on the C3K stellar library (Conroy et al., in prep) to predict the stellar absolute magnitudes in a number of common photometric systems including the \textit{Gaia} DR2 revised passbands.

For the purposes of this work, we restrict our sample to the solar neighborhood and make a magnitude cut to directly compare to \textit{Gaia} DR2 stars with radial velocities. In particular we first select all stars within a 1.0 kpc sphere of the solar position $(-8.2, 0.0, 0.027)$ kpc. \textit{Gaia} DR2 contains the radial velocities for about 7 million stars with $G_{RVS}^{ext} < 12$ -- to mimic this selection, we calculate $G_{RVS}^{ext}$ using the Tycho color transformations presented in \citet{sartoretti2018gaia} and make the same cut. 

For bright stars (most of our sample), the error in the parallax and proper motion can be approximated by a simple error floor of $0.04$ mas and 0.07 mas yr$^{-1}$.  The dependence of radial velocity uncertainty on magnitude is estimated directly from the {\it Gaia} data \citep{katz2018gaia}. The dependence of uncertainties on the \textit{Gaia} scanning law is ignored for this work.  Detailed modeling of the observational uncertainties in the simulations is not necessary for the qualitative comparisons we make in this present study and is therefore left to future work. 

To capture the variance in structure at different azimuths but the same galactocentric radius, we create 18 different solar neighborhoods all separated by 20 degrees but centered at the same galactocentric radius. 

\subsection{Calculating Phase Space Densities}
\label{sec:pd}

A key metric that we will use in the paper is the coarse grained phase space density ($f$) of stars. $f$ is the mass density in a finite six-dimensional volume defined by $d^3x d^3v$, centered on the phase space position of a star at $(x,v)$. Phase space density has proven to be a powerful metric when studying substructure in the halo \citep{helmi2002phase, hoffman2007evolution}, but has not really been used to study structure in the disk. The results presented in this paper utilize EnBID \citep{sharma2006multidimensional}, which builds upon \citet{ascasibar2005numerical}, to numerically calculate the phase space densities. 

At the heart of EnBID's density calculation is a binary space partitioning tree. Instead of assuming {\it a priori} a metric for the multi-dimensional space, EnBID provides a locally adaptive metric using a binary space partitioning tree and an entropy-based splitting criterion. The smoothing for the densities is done using SPH-like kernel-based methods.

\subsection{Variants of the Simulation}
\label{sec:variation}
Many assumptions are made about the Milky Way disk to construct the model described above. The goal of our work is to test these assumptions using a combination of \textit{Gaia} DR2 combined with ground-based spectroscopic surveys. In that vein, we have run three simulations in total:  \\
\begin{itemize}
\item The fiducial simulation with all the model ingredients described above: clustered star formation and realistic disk gravitational potential. 

\item A simulation with an axisymmetric gravitational potential and with clustered star formation. The setup of the axisymmetric simulation is almost identical to the fiducial simulation with one key difference: the non-axisymmetric components of the potential are re-distributed into axisymmetric components. Specifically, the mass of the bar is added to the bulge and the mass in spiral arms is added to the axisymmetric disk. The axisymmetric simulation does not include GMCs for two reasons: (1) the primary role of GMCs in our simulations is that of perturbers,  and (2) star clusters are not born inside GMCs in the simulation. Consequently, the "disruptedness" of star clusters in the axisymmetric simulation depends only on the cluster dissolution model and not on any (large-scale or small-scale) scattering due to the potential. 

\item A simulation with non-axisymmetric perturbations (with bar \& spiral arms) but with no clustered star formation (NCSF simulation hereafter). Instead of forming stars within clusters, we form them as above but in $N=1$ systems. Since there are no clusters to dynamically heat up in this case, the small-scale scattering will have a pretty minor role on the resulting dynamical properties of the disk. Consequently, since running a model with GMCs is five times more computationally expensive than without, we choose not to include them in the NCSF simulation as well.  

\end{itemize}

\section{Global Properties of the Simulated Galaxy}
\label{sec:global}

\subsection{Model Validation}
\label{sec:obs}

In this section we compare several global properties of the simulated galaxy to observations of the Milky Way. Among them, the ones that we consider in this work are: the age-velocity dispersion relation \citep{holmberg2009geneva}, the age-$J_z$ relation \citep{ting2018vertical}, the stellar surface density profile \citep{rix2013milky}, the metallicity distribution function (MDF) in the solar neighborhood \citep{hayden2015chemical}, structure in the U-V plane, and structure in the action-angle plane. In this section we focus on the solar neighborhood sample described in Section 2.7. 

The age-$\sigma_z$ relation provides an important observational constraint, since it has been shown that bluer and therefore younger populations have smaller velocity dispersions in all directions compared to  the redder, older populations \citep{holmberg2009geneva}. The top-left panel of Figure \ref{fig:obs} shows the age-$\sigma_z$ relation for stars in the different solar neighborhoods. The black points show observations from the \citet{holmberg2009geneva}, the grey line shows the input velocity dispersion, and the different lines show the different neighborhoods at the solar radius but different azimuthal angles. The selection function of the Geneva-Copenhagen Survey age-$\sigma_z$ relation is not systematically taken into account but the effect on the immediate solar neighborhood is likely minimal. From this plot, it is clear that the age-$\sigma_z$ relation that is obtained from the fiducial simulation is consistent with what we see in the Galaxy today. 

\begin{figure*}
 \includegraphics[width=168mm]{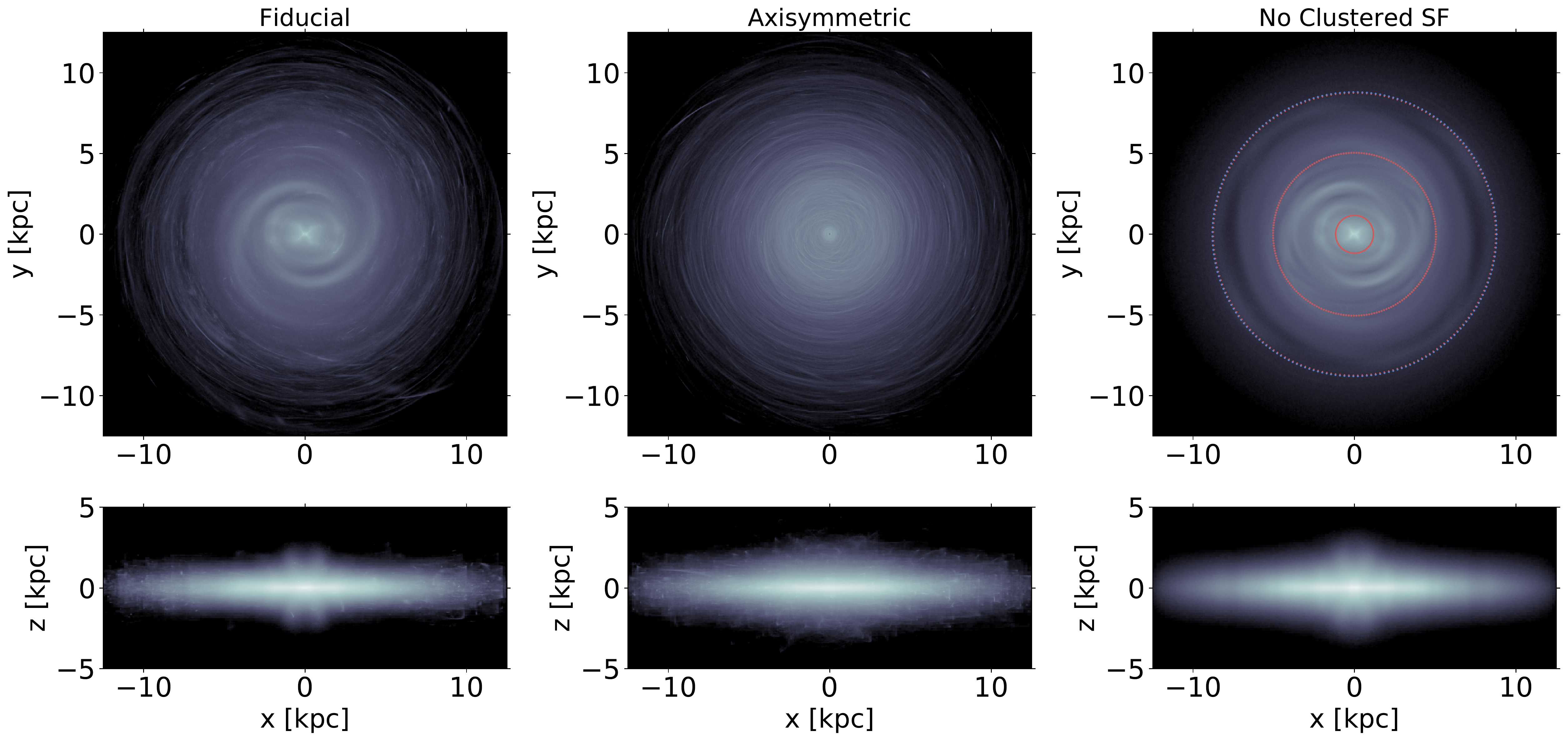}
 \caption{Face-on and edge-on projection of the stellar density in the three different variations of the simulation. \textit{Left panel:} The fiducial simulation, which has both clustered star formation and non-axisymmetries in the potential. We can see both the resonances due to the spiral arms and the bar and individual star clusters being disrupted in the galaxy. \textit{Middle panel:} The axisymmetric simulation, which has clustered star formation but no non-axisymmetries in the potential. The face-on projection is very clearly rich with structure of disrupting star clusters. This simulation presents an extreme control case where structure due to star formation dominates over structure due to non-axisymmetries. \textit{Right panel:} Simulation with no clustered star formation and non-axisymmetries in the potential. This simulation presents a control case where structure due to resonances by non-axisymmetries dominates over structure due to disrupting star clusters. The blue dotted circle shows the corotation radius of the spiral arms in our simulation (approximately at $9$ kpc) and the red dotted circle shows the corotation radius of the bar (approximately at $5$ kpc). The smaller red dotted circles show the inner Lindblad resonance and the outer Lindblad resonance of the bar. The $R_{OLR, bar}$ and $R_{corot, spiral}$ overlap, which has been shown to be a powerful possible mechanism of radial migration.} 
  \label{fig:pretty}
\end{figure*}

A caveat to using the age-$\sigma_z$ relation to assess our GMC scattering model is that heating is often complicated by the overall secular evolution of the disk and radial migration. \cite{ting2018vertical} argue that the age-$J_z$ relation is perhaps more robust, since $J_z$ is an adiabatic invariant under gradual changes of the potential. The top right panel of Figure \ref{fig:obs} shows this plot. The data are the black dots from a \textit{Gaia} DR2 and APOGEE cross-match with the same cuts that were applied to the simulations and the different lines are the different solar neighborhoods. We can see clearly that the slope of the lines and the end-points are not quite the same. Taken at face value, this comparison seems to suggest that the scattering in our models is not quite as strong as data suggests. A possible source of this discrepancy could be underestimating the number of GMCs at $z=0$. The total mass in GMCs in the disk, which controls the number of GMCs that are spawned in the disk, is a somewhat uncertain quantity with some studies \citep{solomon1979giant} implying that the number could be as high as 4000 ($60\%$ greater than our fiducial value). More GMCs would imply a stronger force field and, consequently, stronger scattering.

\begin{figure*}
 \includegraphics[width=168mm]{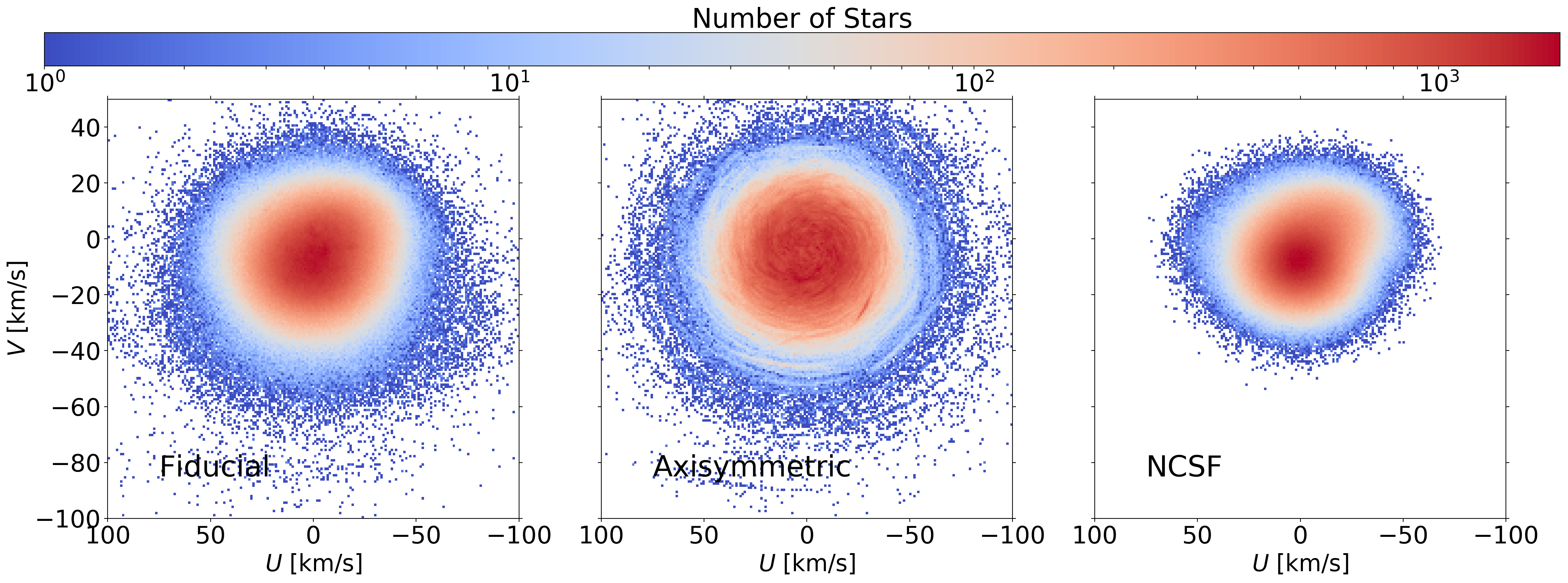}
 \caption{Solar neighborhood UV velocity plane for the three different simulations presented in this work including the fiducial simulation (left), axisymmetric simulation (center), and no clustered star formation (right). The stars shown here are only the ones that are dynamically evolved in the simulations (ages less than 5 Gyr). }
  \label{fig:uv}
\end{figure*}

\begin{figure*}
 \includegraphics[width=168mm]{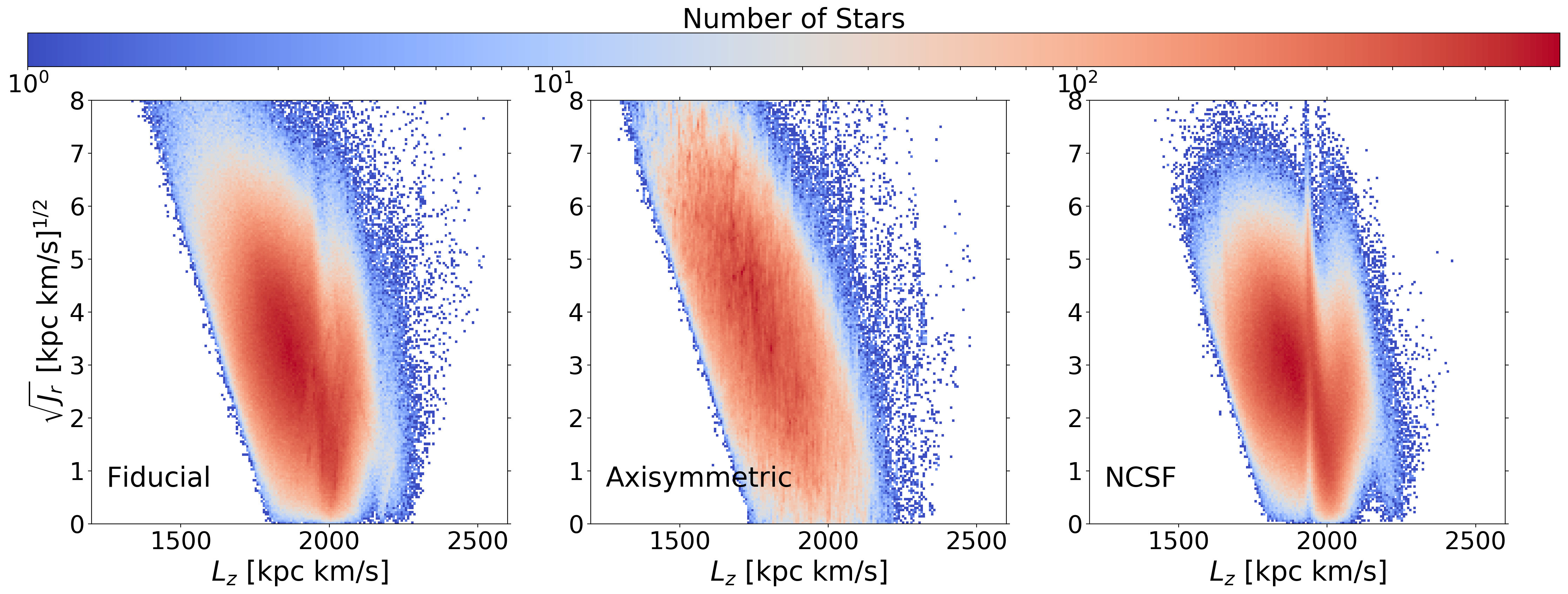}
 \caption{Solar neighborhood action-angle plane for the three different simulations presented in this work including the fiducial simulation (left), axisymmetric simulation (center), and no clustered star formation (right). Recent work has shown rich structure in action-angle space in the solar neighborhood. This set of plots shows the radial oscillations $J_r$ as a function of the angular momentum in the z-direction ($L_z$). The stars shown here are only the ones that are dynamically evolved in the simulations (ages less than 5 Gyr). The fingerprint-like structure in the right panel (NCSF simulation) is likely due to spiral arm and bar resonances -- similar structure is also visible in the fiducial simulation but it is less prominent. }
  \label{fig:action}
\end{figure*}

\begin{figure*}
 \includegraphics[width=168mm]{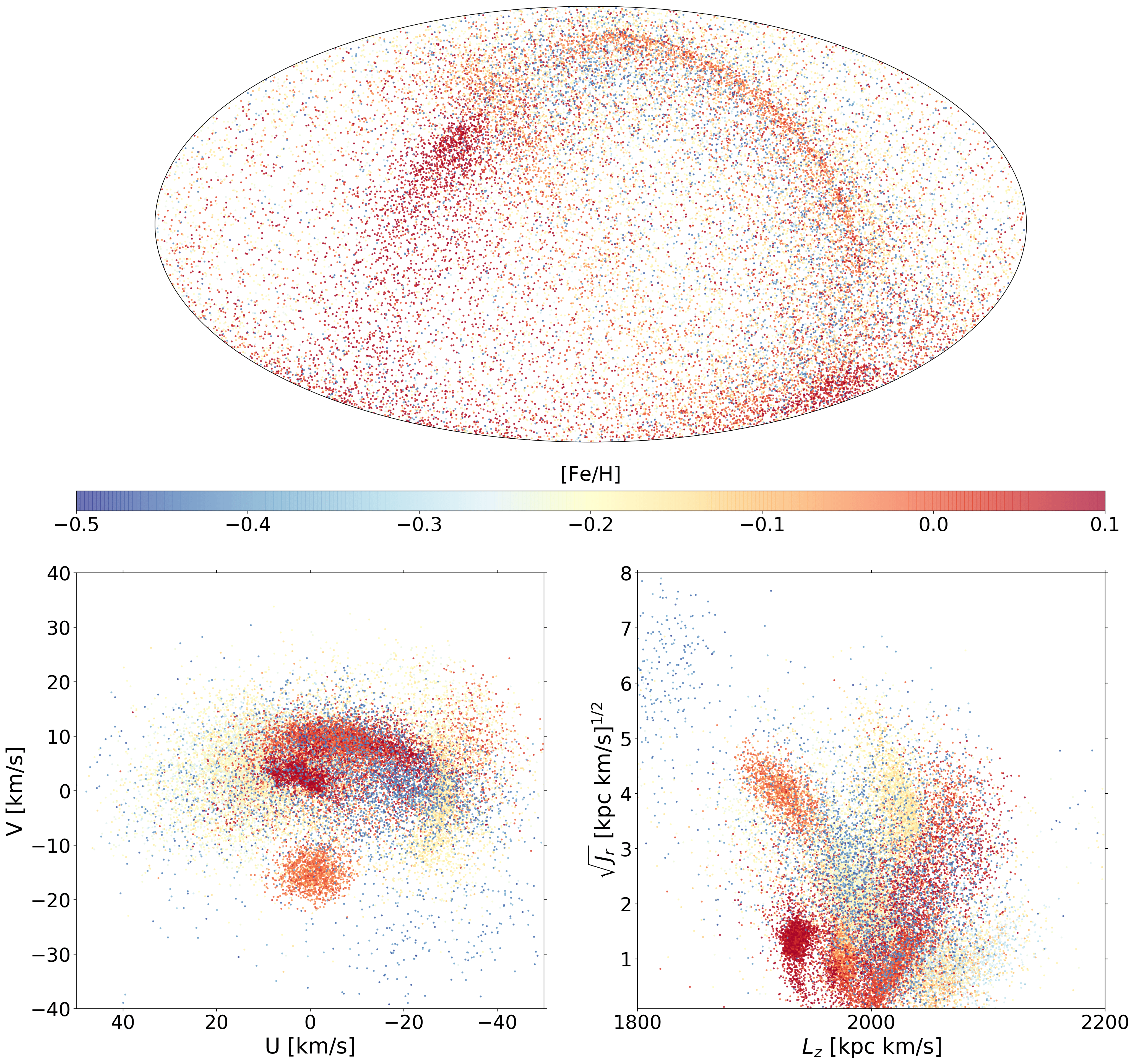}
 \caption{Phase space distribution of stars belonging to the ten clusters that contribute the most stars in the solar neighborhood colored by the birth clusters' metallicities. Top panel: R.A.-Dec. Bottom left: UV plane. Bottom right: $L_z - \sqrt{J_r}$.}
  \label{fig:nclusters}
\end{figure*}

The bottom left panel shows the stellar mass surface density profile of the simulation and an exponential profile with a scale length of $R_s = 2.2$ kpc \citep{rix2013milky} calibrated to a solar stellar surface density value of $38 M_{\odot}$ pc$^{-2}$. The light grey highlighted regions reflect the uncertainty in the value of $R_{\odot}$ and $\Sigma(R_{\odot})$ \citep{bland2016galaxy}. As is clear from the plot, the surface density profile of the simulation at the solar radius agrees well with an exponential profile with a $R_s = 2.2$ kpc scale length. This observational test reaffirms that the SFH of the simulation, the potential of the simulation, and the background model are broadly consistent with that of the Galaxy. 

The bottom right panel shows the comparison of the metallicities of stars that end up in the solar neighborhoods in the simulation with APOGEE stars in the solar neighborhood. The catalog used for the metallicity values for APOGEE is from \citep{ting2018payne} and the same cuts applied in \citet{hayden2015chemical} were applied here (high SNR, cool, giant stars). We see an offset in the simulation metallicities of about ~-0.1 dex at higher metallicites; i.e., stars in the simulation at the solar radius are more metal-poor than expected. The most likely explanation for this discrepancy is weaker radial migration than expected \citep{minchev2013chemodynamical}. Radial migration leads to inner-Galaxy, metal-rich stars migrating outwards due to a variety of dynamical effects discussed earlier. In the case of our simulation, we do not have transient spiral arms, which are much more efficent at radial migration. More quantitatively, for stars about $5$ Gyr old in our simulations, we find that the $<R - R_{birth}>$ for stars that end up in the solar neighborhood is about $1.5$ kpc. \citet{frankel2018measuring} argue for a higher radial migration of roughly $2.8$ kpc for stars of a similar age (Eq. 28). However, the standard deviation of the simulation [Fe/H] and the APOGEE metallicities is almost identical at ~0.24 dex and ~0.25 dex respectively. We expect that the modest 0.1 dex offset to have little effect on the main conclusions and prediction presented in this work.

\subsection{Phase Space Signatures of Clustered Star Formation} 
\label{sec:pd_csf}

In this section we discuss the phase space signatures of clustered star formation. In Figure \ref{fig:pretty} we provide an overview of the structure in the three different simulations described in Section \ref{sec:variation} in the x-y and x-z projection. The left panel shows the fiducial simulation that includes all the model recipes described in Section 2, the middle panel shows the axisymmetric potential simulation, and the right panel shows the no clustered star formation simulation.  There are very clear differences between the three models.

\begin{figure}
 \includegraphics[width=84mm]{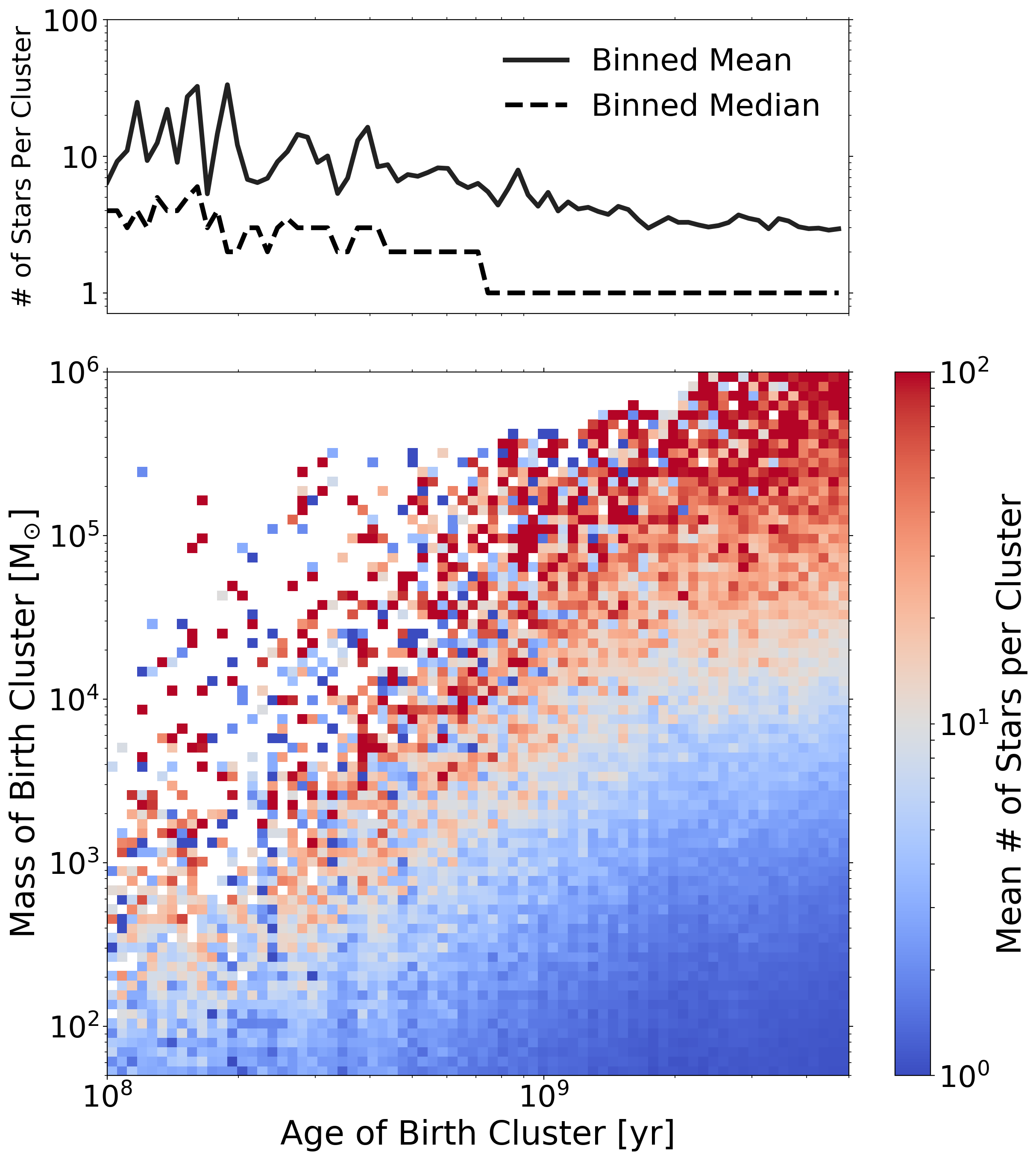}
 \caption{Number of stars that clusters of different masses and ages contribute to a \textit{Gaia}-like solar neighborhood in the simulation at $z=0$. Top Panel: Mean number of stars per cluster in the solar neighborhood sample of the simulation (within $1.0$ kpc of the solar position) as a function of age of the cluster. Bottom Panel: Mean number of stars per cluster in the solar neighborhood sample of the simulation binned in both age of the birth cluster and the mass of the birth cluster.}
  \label{fig:demo}
\end{figure}

\begin{figure*}
 \includegraphics[width=168mm]{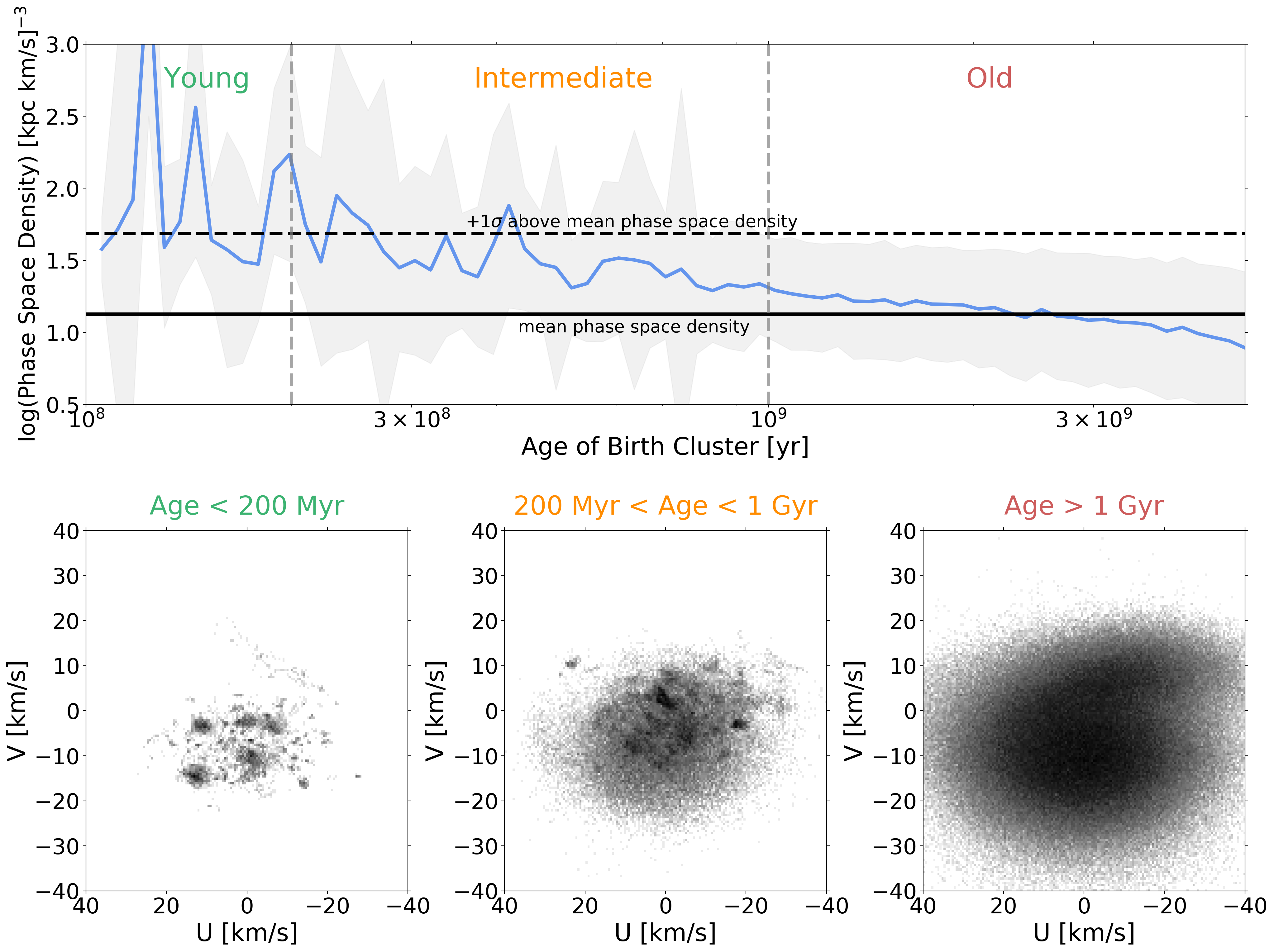}
 \caption{Top panel: Mean phase space density as a function of age.  Stars with ages less than $200$ Myr clearly lie in overdense regions of the phase space density distribution. There are a large number of stars in the age range $200 <$ age $< 1000$ Myr that are in slightly overdense regions of the phase space density distribution but not above the overall mean plus $1 \sigma$ phase space density implying that they would not be as easily "visible" as younger stars. Stars older than $1$ Gyr are mostly phase-mixed with the background population of field stars. Bottom panels: UV diagrams for stars in three age bins.  The amount of substructure is clearly very sensitive to the age of the population.}
  \label{fig:visibility}
\end{figure*}

As one would expect, the axisymmetric simulation (center panel) shows many disrupting star clusters scattered all throughout the galaxy. Since this simulation does not have any large-scale scattering processes influencing individual stars, a lot of the structure due to star formation we see is more intact than one would expect in a realistic galaxy. Instead, the spread that we see in the distribution of the stars from the same cluster is solely through the cluster disruption recipe described in Section 2. The axisymmetric simulation provides an extreme limiting case for the amount of small-scale structure expected in the galactic disk. In the absence of non-axisymmetric components, the original stellar associations are readily identifiable simply from their positions in the configuration space

The simulation with no clustered star formation (right panel) is the limiting case where every star is born by itself, in a galaxy with two strong non-axisymmetric components. Consequently, any clumping or noticeable structure in phase space is solely due to dynamical resonances and not to clustered star formation. Much of the smaller scale structure we saw in the center panel is completely smoothed out. The corotation radii for the spiral arms and the bar and the innner/outer Lindblad resonances  \citep{binney2011galactic} for the bar are overplotted as well. The OLR of the bar and the corotation radius of the spiral arms overlap only $600$ pc away from the solar position for the values we have chosen for the spiral arm and the bar pattern speed. Some have argued that the local velocity substructure in the solar neighborhood could be caused by the resonance overlaps of the spiral arms and bar \citep{monari2016effects}. 

Finally, the fiducial simulation is shown in the left panel. It is readily apparent that there are spiral arms and a bar in this simulation. However, we can also see thin banana-like overdensities that correspond to dissolving star clusters. The most important point to take away from this Figure is that we know that the Milky Way lies somewhere in the middle -- we know that the Galaxy has clustered star formation and that the Galaxy has a bar and spiral arms. The ultimate goal of this project is to provide testable observational signatures that can help us get closer to recovering the history of the former using kinematics, chemistry, or both.

A more detailed comparison of the three model variants is presented in Figures \ref{fig:uv} and \ref{fig:action}. Figure \ref{fig:uv} explores the UV plane in the simulations, where U is the radial velocity positive in the direction of the galactic center and V is the tangential velocity in the direction of the galactic rotation. We focus on ages < 5 Gyr as those are the stars that are dynamically evolved in the simulation. On the left is the fiducial simulation, in the middle is the axisymmetric simulation, and on the right is the NCSF simulation. The UV plane has a long history of being employed to find cold structures in phase space and has yielded considerable success \citep[e.g.,][]{dehnen1998distribution}. The middle panel shows the structure in the UV plane when there is no large-scale or small-scale scattering due to the potential -- in this case, there is clearly more structure preserved in phase space about clustered star formation. The right panel shows the structure in the UV plane when there is no clustered star formation and we can see no small overdensities but large-scale  structure is visible. The fiducial simulation on the left panel has both large-scale and small-scale structure due to scattering by the potential and clustered star formation. Disentangling phase space structure at these two different scales motivates the simulations presented in this work. 

Figure \ref{fig:action} shows the radial oscillations, $J_r$ as a function of $L_z$ for the simulations. Similar to the previous plot, only dynamically evolved stars in the simulation (age < 5 Gyr) are included. Recent work \citep[e.g.,][]{trick2018galactic, sellwood2018discriminating} has shown the power of the action-angle framework in decomposing different dynamical signatures of the various components of the galaxy. The persistence of a strong vertical fingerprint in the NCSF simulation points to the resonance overlap mentioned above and similar signatures have been seen in different simulations \citep{sellwood2018discriminating}. Overall, we find similar structure as previous work but not quite as detailed and rich, likely due to the absence of complexity in our spiral arms and bar model and not evolving the full galaxy for $13$ Gyr. 

Figure \ref{fig:nclusters} picks out the ten clusters that contribute the most number of stars to the solar neighborhood today and shows them in the R.A.-Dec. plane, the UV plane, and the action-angle plane. Stars are color-coded by their metallicity. The top panel of this figure shows us the potential structure that dissolving star clusters could imprint on configuration space. This kind of structure would be very difficult to find given the dominant background stars and the scales at which stars are spread out (however, see \citealt{meingast2019extended} for recent examples). However, the bottom two panels show how phase space holds more information about co-natal stars both in the velocity plane and the action-angle plane. Moreover, the color-coded chemistry of individual star clusters also shows how chemistry could potentially add even more information. The clusters shown in this Figure have masses anywhere from $ \sim 10^3 - 10^6$ $M_{\odot}$ and contribute anywhere from 1500 to 4000 stars. 

\section{Identifying Disrupted Clusters with Kinematics and Chemistry}
\label{sec:kinematic_chemical}

\begin{figure*}
 \includegraphics[width=168mm]{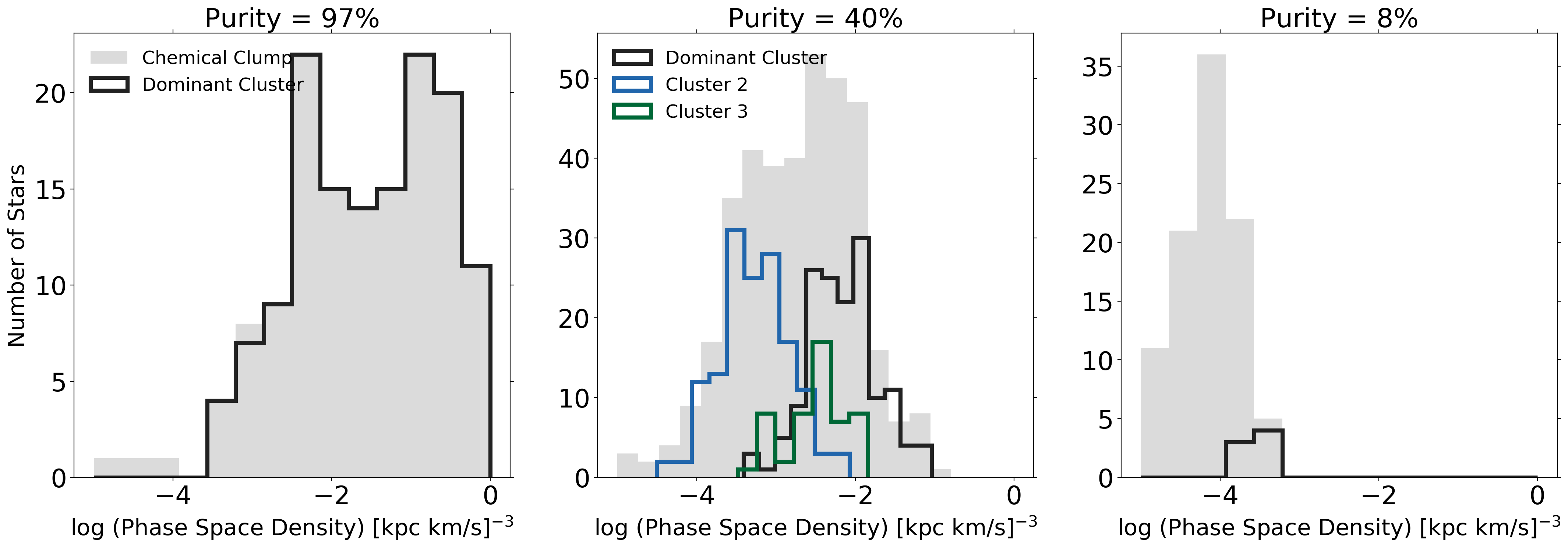}
 \caption{Distribution of phase space densities for three different chemical clumps found by HDBSCAN of varying purities. In each panel we show the phase space density distribution of all stars in the clump (grey) and the one or more most significant clusters that comprise the clump (solid lines). \textit{Left:} The majority of stars found by the clustering algorithm for this clump in solely chemical space are actually co-natal. As expected, the phase space density of all stars in the clump tracks the phase space density of the dominant star cluster. \textit{Center:} The chemical clump found here has a purity of around $40\%$. Two star clusters with very similar chemistries contribute an equal number of stars to this impure chemical clump and one minor star cluster and a few background stars make up the rest of the chemical clump. The difference in phase space densities for the two dominant clusters is readily apparent and shows the potential power of combining chemical and kinetic information. \textit{Right:} Most stars in this chemical clump are field stars. The phase space density here does not help discriminate between the field population and the few stars that are actually co-natal. }
  \label{fig:chemical_clump}
\end{figure*}

\subsection{Demographics}

We now move on to studying the behavior of the simulations on smaller physical scales. One of the first questions that we can directly ask with the simulations is what are the demographics of stars present in the solar neighborhood today? The top panel of Figure \ref{fig:demo} shows the mean and median number of stars that every cluster contributes to the solar neighborhood at $z=0$ as a function of the age of the birth cluster. The bottom panel of Figure \ref{fig:demo} shows the 2D histogram of the mean number of stars per cluster in the solar neighborhood today as a function of cluster mass and age. As one would expect, younger star clusters contribute more stars because they are likely less phase-mixed than older star clusters. In the lower panel, for the average cluster mass of $500 M_{\odot}$, we can notice from the plot that for ages above 1 Gyr, they only contribute $\sim1$ star to the solar neighborhood on average but below 1 Gyr, we are sampling more stars from clusters of this mass implying that there could be some discernible phase space signature of dissolving clusters in this mass bin. Figure \ref{fig:demo} also shows that assuming stars > 5 Gyr (background stars) to be completely phase-mixed is a reasonable assumption. 

However, Figure \ref{fig:demo} does not tell us about whether stars born together are actually close together in phase space. Consequently, we would like to ask the question: how long does a cluster remain discernable in phase space?  Figure \ref{fig:visibility} shows the binned mean phase space density  of stars as a function of their age. We can see that young stars ($< 2 \times 10^8$ yr) reside in overdense regions in phase space. This is not unexpected as a lot of this structure is dominated by star clusters in early stages of their disruption. More interestingly, there is a large age range where stars are above the mean phase space density but only marginally ($2\times10^8$ yr < age $<  10^9$ yr), and beyond that, most stars are at or below the mean phase space density implying that they are well mixed with the field population. The second regime, where there is some information but not quite enough to pick out individual star clusters, is the more interesting one, since many stars fall in that regime. The question that we will focus on for the rest of the paper is how to find stars that were born together that have some information content in phase space but not enough for them to be trivially identified as bound clusters. 

The bottom panels of Figure \ref{fig:visibility} show the UV plane for the different regimes discussed above. As expected, there is rich structure in phase space for the very young stars (bottom left panel) and no discernible phase space signature for phase mixed stars (bottom right panel). The presence of some discernible structure in the bottom middle panel shows the potential of being able to use phase space information to probe clustered star formation. 

\begin{figure}
 \includegraphics[width=84mm]{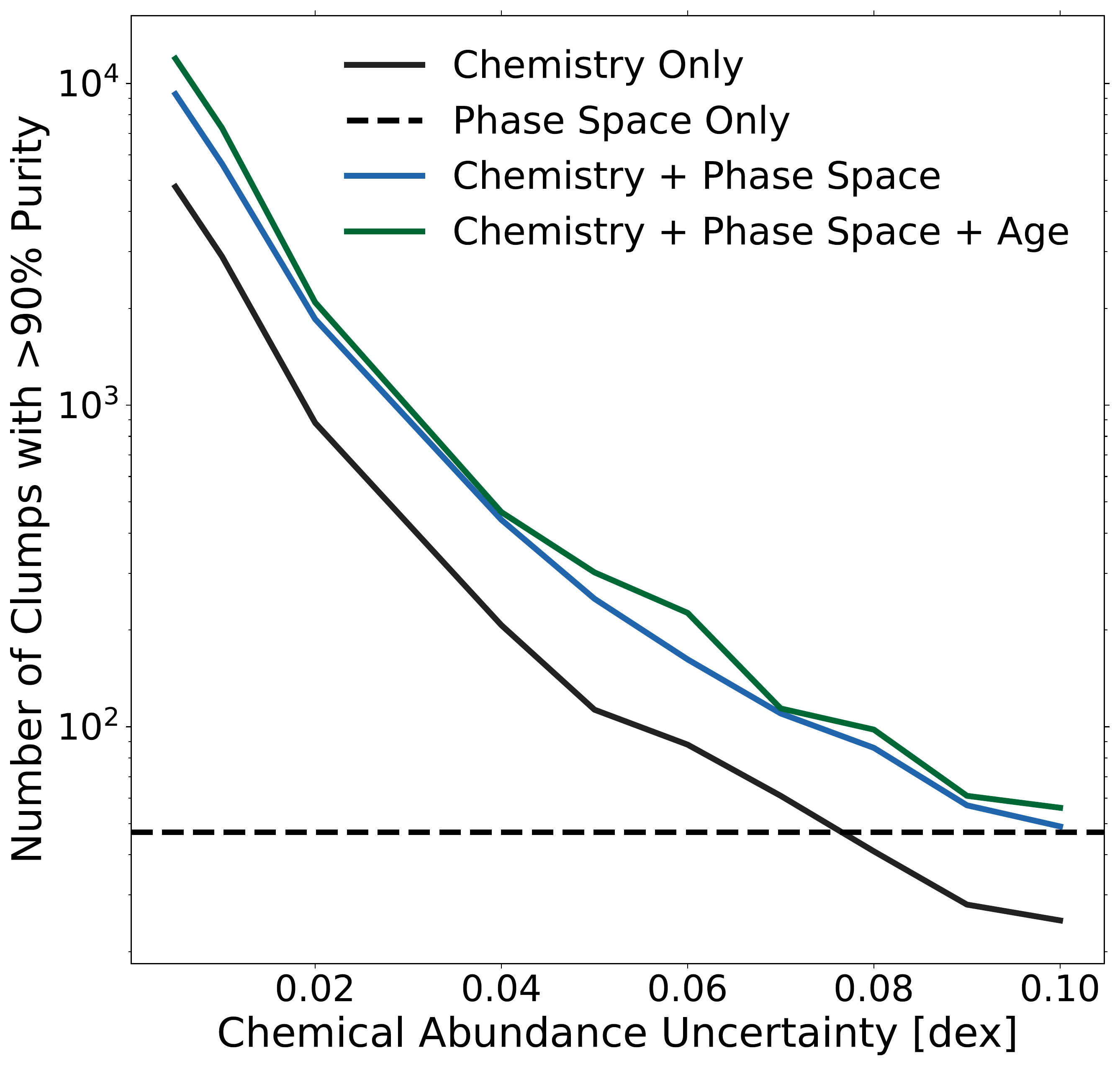}
 \caption{Number of high-purity clumps as a function of element abundance uncertainty when using different combinations of phase space, chemistry, and age information. The black line shows the number of pure clumps found only in chemical space as a function of abundance uncertainty. The black dashed line shows the number of clumps found solely in phase space. The blue and the green line show the relative gain in the number of clumps when phase space and age information are added. }
  \label{fig:chemtag_ps}
\end{figure}

\begin{figure}
 \includegraphics[width=84mm]{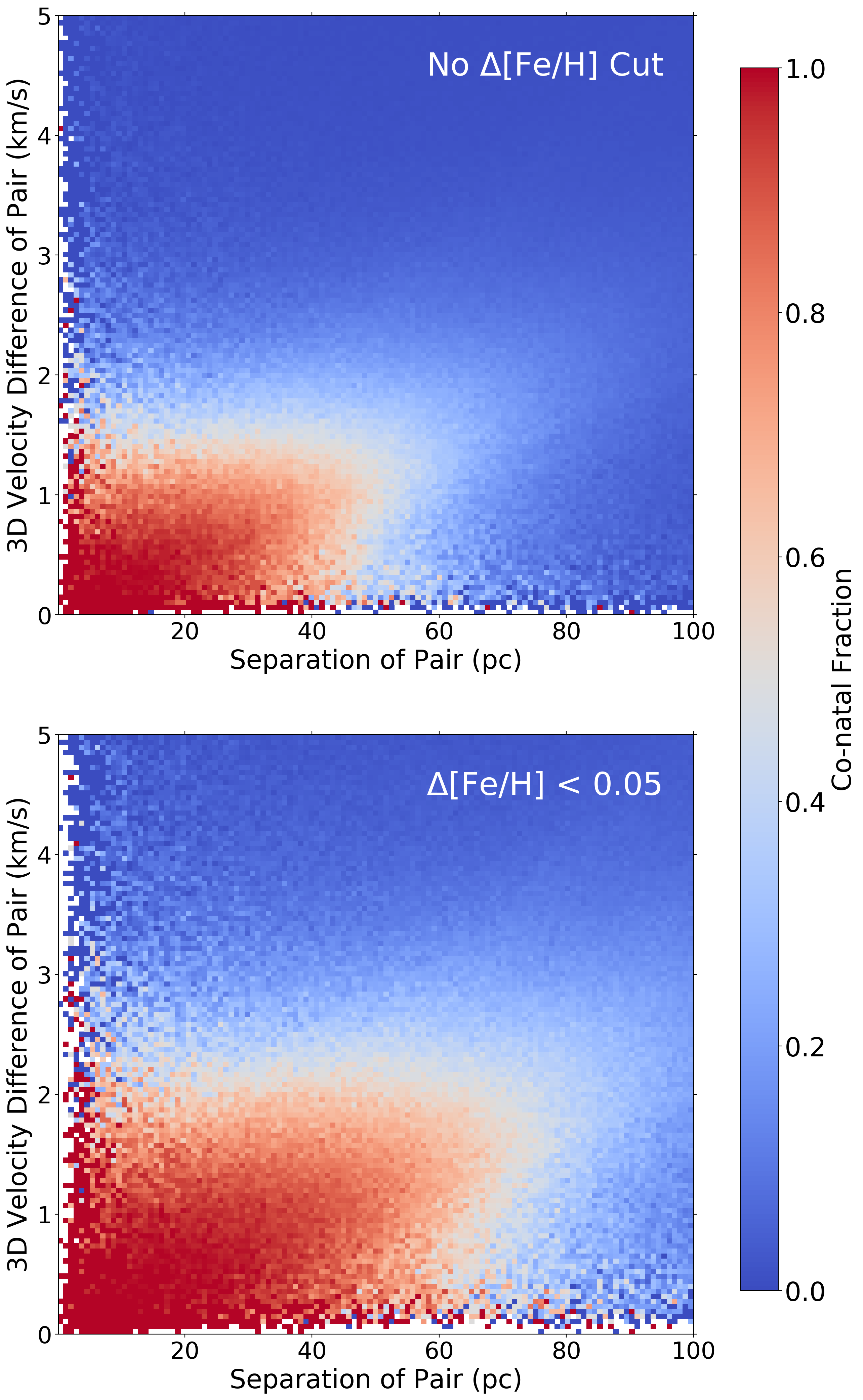}
 \caption{Co-natal fraction of pairs as a function of separation and relative velocity difference. The top panel shows all co-moving pairs with no metallicity information included and the bottom panel shows co-moving pairs with a metallicity difference of $<0.05$ dex. }
  \label{fig:conatal}
\end{figure}

\begin{figure}
 \includegraphics[width=84mm]{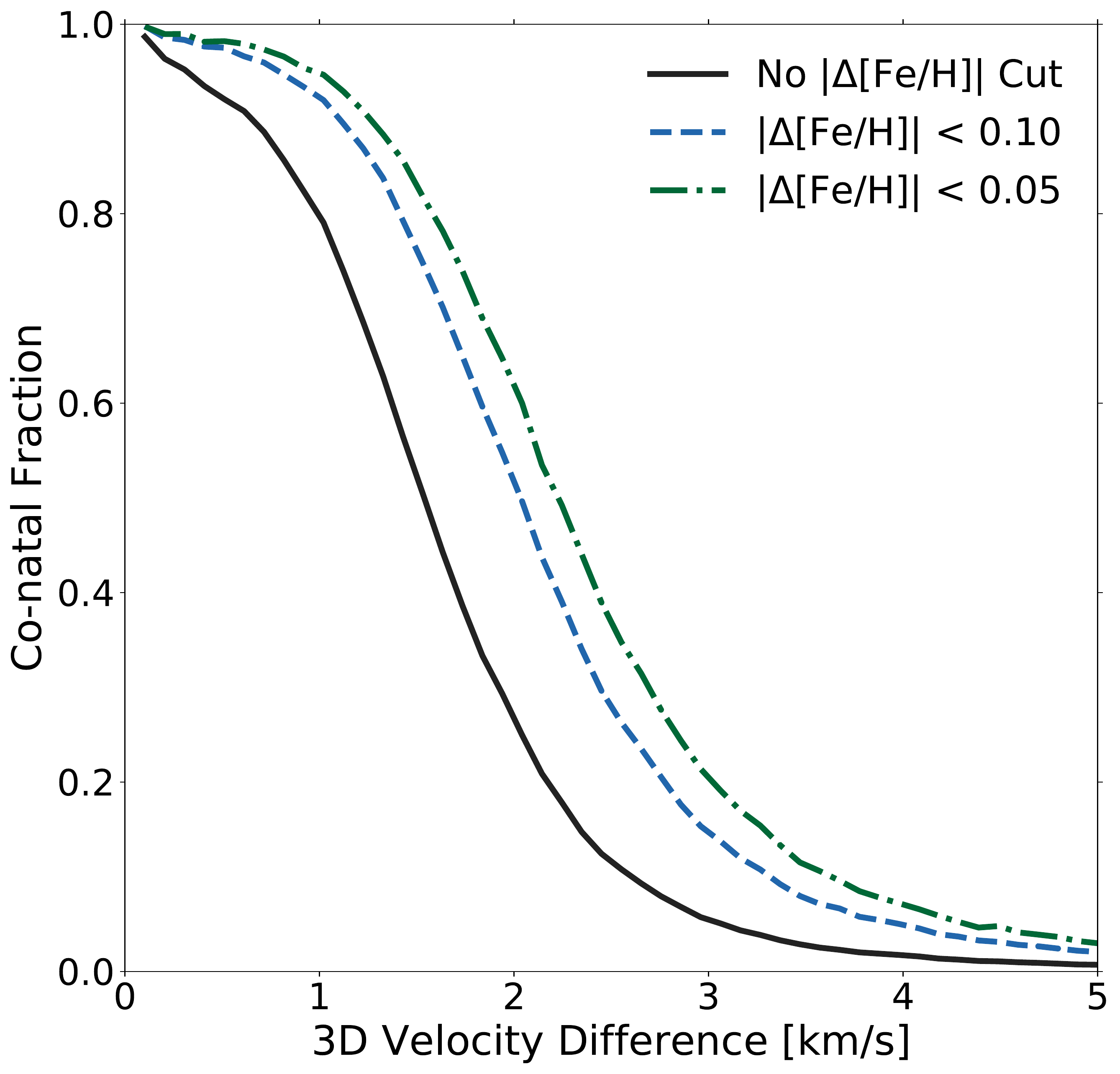}
 \caption{Fraction of pairs that are co-natal as a function of 3D velocity difference. Pairs are selected within a 3D physical separation of $<20$ pc. Results are shown both without a metallicity separation cut, and with metallicity separations of $<0.05$ and $<0.10$.}
  \label{fig:conatal_cut}
\end{figure}

\subsection{Structure in Kinematic \& Chemical Space}

In this section we explore the benefits of combining the information in phase space and chemical space in order to study clustered star formation. To quantify the structure in chemical and phase space in detail, we utilize Hierarchical Density-Based Spatial Clustering of Applications with Noise (HDBSCAN) \citep{campello2013density, mcinnes2017hdbscan}. HDBSCAN builds upon DBSCAN, which is a popular density-based clustering algorithm that uses a parameter called epsilon to determine some density threshold over which to find clumps. HDBSCAN enhances this algorithm by performing DBSCAN over various epsilon values and integrates the result to find a clustering that gives the best stability over different epsilon values (where stability refers to the persistence of the same clusters at different values of epsilon). The real power of this algorithm is that it allows us to find clusters of varying densities (unlike DBSCAN). Both of these properties make the algorithm ideal to use for the problem of chemical tagging since we expect chemical clumps of varying densities and we want to minimize the number of spurious clumps. 

Since our chemical model is semi-empirically derived from APOGEE, the dimensionality of the chemical space is 19. Figure \ref{fig:chemical_clump} shows three chemical clumps of varying degrees of purity found by HDBSCAN in the chemical space spanned by the simulation. The purity of a clump is the number of stars from the dominant birth cluster in a clump found by HDBSCAN divided by the total number of stars in that clump. A purity of $50\%$ would imply that half the stars in an HDBSCAN chemical clump were born together in the same cluster and the other half are contaminating field stars. In each panel, the phase space density of the entire chemical clump is plotted in grey and the constituent star clusters' phase space densities are plotted in different colors. The chemical clump on the left is very pure -- almost all stars found to be similar in chemical space were indeed born together. As one would expect in this scenario, the phase space density of both the entire chemical clump and that of the dominant star cluster is almost identical. 

The middle panel shows a chemical clump with a purity of 40\%. Two star clusters born at different times and in different parts of the galaxy contribute roughly the same number of stars to this chemical clump and make up about 80\% of the stars belonging to this clump. The other 20\% are from another star cluster and background stars. The phase space density distribution of the star clusters shows how one could use phase space information to break the degeneracy in impure chemical clumps. 

Finally, the right panel shows a chemical clump that is dominated by field stars that happen to have a similar chemical composition. In this case, the phase space density does not seem to be a useful metric in breaking the degeneracy between co-natal and field populations. 

Figure \ref{fig:chemtag_ps} shows the number of clumps with $90\%$ purity found by HDBSCAN as a function of abundance uncertainty in just chemical space, just phase space, chemical space and phase space combined, and chemical space, phase space, and age combined. At almost all abundance uncertainties, the number of clumps found with the addition of phase space information to chemical space is double that found in chemical space alone. The ages are given an observationally-motivated uncertainty of 0.3 dex and for this reason do not significantly increase the total number of pure clumps identified. Moreover, it is worthwhile to note that the abundance precision should be regarded as "effective" precision, since we do not take into account the co-variances. That is, a nominal 0.05 dex precision could have a much smaller "effective" uncertainty due to the co-variances \citep{ting2015apogee}. Moreover, the number of independent dimensions in our chemical model (and APOGEE) is difficult to measure -- a chemical model with a tunable parameter for the number of independent dimensions in chemical space is subject of future work.  

\subsection{Co-moving Pairs}

Recently, there has been interest in the study of pairs of stars that are close together and moving at similar velocities, so-called co-moving pairs \citep{oh2017comoving, andrews2017wide, gagne2018banyan, price2017spectroscopic, andrews2018using}. \citet{oh2017comoving} found a suprisingly large number of co-moving pairs with separations of $\geq 1$ pc and small relative velocities. There are two possible origins for these pairs: either they were born together and are moving through the Galaxy together ("co-natal") or they happened to be co-moving either randomly or due to resonances by external perturbers such as the bar or spiral arms. 

The simulations presented in this work offer a useful testbed to investigate the nature of these $N=2$ clumps in the disk. Figure \ref{fig:conatal} shows the co-natal fraction, the number of pairs in a bin that were born together divided by the total number of pairs in that bin, plotted as a function of the physical separation and the 3D velocity difference. The top panel shows all the co-moving pairs with no metallicity information included and the bottom panel shows all the co-moving pairs with a metallicity difference of $<0.05$ dex. The top panel is particularly striking because it suggests that stars currently separated by as much as $20$ pc have a  high co-natal fraction for $\Delta V<2$ km s$^{-1}$. Moreover, adding chemical information allows us to find co-natal stars that are even further apart and separated by a higher relative velocity. 

Figure \ref{fig:conatal_cut} shows the co-natal fraction for all pairs within $20$ pc as a function of $\Delta V$.  We include cases with no metallicity information (solid line) and cases with a relative metallicity difference of $<0.05$ (dot-dashed) and $0.10$ (dashed).  Inclusion of even a modest amount of metallicity information results in much higher co-natal fractions out to several km s$^{-1}$. The results presented here are promising for Galactic archaeology and the quest of finding stars born together.

\section{Caveats, Limitations, \& Future Work}
\label{sec:discussion}
In this section we discuss the limitations of the model. We have taken the approach in this work to develop a complete dynamical model of all stars born in the Galactic disk over the past 5 Gyr. As a result, a number of assumptions and approximations were made. Below, they are listed in order of roughly decreasing importance. Many of these will be tested and explored in future work. 

\begin{enumerate}

\item Perhaps the largest uncertainty in our model is how star clusters are born and disrupted in the galaxy. The uncertainty stems from ongoing debates in the star cluster community regarding the interplay between and the relative importance of individual clusters becoming unbound due to gas expulsion ("infant mortality") \citep{lada2003embedded} versus the hierarchical structure scenario \citep[e.g.,][]{kruijssen2012fraction}. Some other formation channels \citep{longmore2014formation, krumholz2018star} have been put forth as well. One of the main goals of this broader project is to identify features in phase space that might be able to shed new light on this debate. We plan to run simulations with different star cluster birth and disruption recipes and hope to place some constraints on those by comparing their phase space signatures to \textit{Gaia} DR2. 

\item A limitation of the current model is that of a mostly steady-state potential. The spiral arm pattern and the bar are evolving with time but at a fixed pattern speed and a prescribed mass. However, we expect that these non-axisymmetric components might be transient features. Furthermore, the radial migration in our simulation, as implied in Figure \ref{fig:obs}, is not quite as strong as expected and, consequently, the metallicity distribution functions are offset at high metallicities from their expected value. On the other hand, not much is known about the time evolution of the spiral arms, bar, and GMCs, which poses a difficult modelling challenge. 

\item An area for future improvement is the treatment of \textit{N}-body interactions within star clusters via test particle integration. In this work we adopted a simple scheme of undervirializing star clusters to mimic the effect of self-gravity at early times. However, we expect this to depend on the mass, density, and galactocentric location of star clusters in the galaxy. In future work we plan to run a grid of simulations to create an undervirialization recipe that takes into account the factors mentioned above. 

\item There are several assumptions made in the chemical model that should be more thoroughly explored in future work. The spatiotemporal correlations between the chemical signature of clusters born near each other at similar times \citep{krumholz2017metallicity} is mostly ignored but could play an important role in the feasibility of chemical tagging in the galaxy. Moreover, the covariances between the individual abundances are largely ignored -- however, we know this to be incorrect in detail \citep{ting2015apogee}. The dimensionality of chemical space is also not known. Future iterations of the chemical model will attempt to take these factors into account in a principled way by modeling the covariances between measured chemical abundances and including the number of indpendent dimensions in chemical space as a tunable parameter. 

\item GMCs are likely an important component of the vertical heating of the Milky Way disk and also play a role in radial diffusion. There are three main sources of uncertainty in our GMC model: how many GMCs there are, how large they are, and how to efficiently calculate their force at a given point at reasonable computational expense. All of these, separately and/or combined, play an important role in the radial and vertical heating of the simulated galaxy. A more comprehensive picture for GMC formation and evolution such as \citep{jeffreson2018general} could be explored.

\item The radial growth for the simulated galaxy that we adopt is based on a fairly simple relation derived from extragalactic data \citep{schruba2011molecular}. It is straightforward to consider alternative models for the global evolution of the disk structure, and we could for example adopt relations from cosmological zoom-in simulations \citep[e.g.,][]{sanderson2018synthetic}.

\item The CMF, in particular its higher mass cutoff and possible redshift evolution, is quite uncertain.  The CMF is important because it determines what kind of clusters end up in the solar neighborhood and the overall clumpiness of the chemical space. Exploring the CMF high-mass cutoff is planned for a future work. 
\end{enumerate}

All of the assumptions and limitations discussed above present fruitful future research directions, especially in light of \textit{Gaia} and present and upcoming ground-based spectroscopic surveys. 

\section{Summary}
\label{sec:summary}

In this paper we have introduced a new set of simulations that are the first of their kind to model the full population of stars (younger than 5 Gyr) comprising a Milky Way-like disk. These simulations allow us to resolve the small-scale structure in phase space and chemical space due to clustered star formation. Moreover, we have run two additional control simulations in order to isolate the effects due to clustered star formation and resonances. 

We present a simple model for how to mimic star cluster initialization, evolution, and disruption based on a combination of analytic theory, hydrodynamical+\textit{N}-body simulations, and observations. The model is calibrated to broadly agree with the cluster formation efficiency of a Milky Way-like galaxy and the evolution of the mass-radius relation of various cluster formation channels.

The star cluster model is coupled to a realistic model for the potential of the Milky Way that includes a bar, spiral arms, GMCs, and an axisymmetric component. We assume that all stars are born in star clusters, and evolve stars in the mass range $[0.5, 1.5]\, M_{\odot}$ born in the last 5 Gyr ($\sim4$ billion stars in total) based on the initial conditions described above. 

Our results are summarized as follows:

\begin{itemize}

\item The simulations of the Milky Way disk presented in this work agree with observations of the age-velocity dispersion relation, the surface density profile, and the metallicity distribution function in the solar neighborhood of the Milky Way. 

\item The phase space signatures of clustered star formation are, as expected, a strong function of age. We find a large range of ages where structure in phase space should still be visible above the background of old, phase-mixed clusters. Moreover, the simulations presented in this work present a unique opportunity to study the demographics of the birth sites of stars in the solar neighborhood today. 

\item Similar to previous work \citep[e.g.,][]{ting2015prospects}, we find that solely looking in the chemical abundance space leads to impure chemical clumps, which poses problems for strong chemical tagging. We find that independent of abundance measurement uncertainty, the addition of phase space information leads to an increase of more than a factor of two in the number of pure clumps ($>90\%$ purity) found by HDBSCAN. These results bode well for harnessing the synergies between spectroscopic surveys and \textit{Gaia}. 

\item A high fraction (from $30$ to $70\%$) of co-moving pairs with a large separation (extending to $\sim40$ pc) and a low relative velocity ($<2$ km s$^{-1}$) were born together ("co-natal") in the simulation. Adding a simple $\Delta$ [Fe/H] cut further increases the co-natal fraction at higher separations and velocities. 

\end{itemize}

The models presented in this work provide a unique opportunity to study the imprints of clustered star formation on the kinematics and chemistry of stars in the \textit{Gaia} era. In the near future, we will explore co-moving pairs in \textit{Gaia} DR2 and investigate the predictions of whether high-separation and low-velocity pairs are truly co-natal (Kamdar et al., in prep). Moreover, we will also explore alternate models for clustered star formation and the potential and its evolution with the goal of using \textit{Gaia} data to discriminate amongst various model options. 

\acknowledgements
The computations in this paper were run on the Odyssey cluster supported by the FAS Division of Science, Research Computing Group at Harvard University. HMK  acknowledges  support  from  the  DOE  CSGF  under  grant  number DE-FG02-97ER25308. CC acknowledges support from the Packard Foundation. YST  is  supported  by  the NASA Hubble Fellowship grant HST-HF2-51425.001  awarded by the Space Telescope Science Institute.

This work has made use of data from the European Space Agency mission
\textit{Gaia} (\url{https://www.cosmos.esa.int/gaia}), processed by the Gaia Data
Processing and Analysis Consortium (DPAC,
\url{https://www.cosmos.esa.int/web/gaia/dpac/consortium}). Funding for the DPAC
has been provided by national institutions, in particular the institutions
participating in the Gaia Multilateral Agreement. The Sloan Digital Sky Survey IV is funded by the Alfred P.Sloan Foundation, the U.S. Department of Energy Office ofScience, and the Participating Institutions and acknowledgessupport and resources from the Center for High-Performance Computing at the University of Utah. 

{\it Software:} \texttt{IPython} \citep{perez2007ipython}, \texttt{Cython} \citep{behnel2010cython}, \texttt{Numba} \citep{lam2015numba}, \texttt{Schwimmbad} \citep{price2017schwimmbad}, \texttt{Astropy} \citep{2013A&A...558A..33A, 2018AJ....156..123A}, \texttt{NumPy} \citep{van2011numpy}, \texttt{SciPy} \citep{scipy}, \texttt{AMUSE} \citep{portegies2011amuse}, \texttt{HDBSCAN} \citep{mcinnes2017hdbscan}, \texttt{scikit-Learn} \citep{pedregosa2011scikit}, \texttt{AGAMA} \citep{vasiliev2018agama}, \texttt{Matplotlib} \citep{hunter2007matplotlib}, \texttt{Pandas} \citep{mckinney2010data}, \texttt{Seaborn} \citep{michael_waskom_2014_12710}, \texttt{yt} \citep{2011ApJS..192....9T}, and \texttt{ACRONYM} (https://github.com/bacook17/acronym)

\bibliographystyle{yahapj}

\bibliography{main}

\end{document}